\documentclass{jfm}
\usepackage{graphicx}
\usepackage[greek,english]{babel}
\usepackage{amsmath,bm,mathrsfs}
\usepackage{epstopdf, epsfig}
\usepackage{color}

\newcommand{\gpi}{\textrm{\greektext p}}
\newcommand{\myadd}[1]{{\textcolor{black}{#1}}}

\shorttitle{Impact of non-frozen turbulence}
\shortauthor{H. Tian and B. Lyu}

\title{The impact of non-frozen turbulence on the modelling of the noise from serrated trailing edges}

\author{H. Tian\aff{1}
 \and B. Lyu\aff{1}\aff{2}\corresp{\email{b.lyu@pku.edu.cn}}}

\affiliation{\aff{1}State Key Laboratory of Turbulence and Complex Systems, College of Engineering, Peking University, No.5 Yiheyuan Road, Haidian District, Beijing 100871, PR China
\aff{2}Laoshan Laboratory, Qingdao 266100, PR China}

\begin{document}

\maketitle

\begin{abstract}
	
 Serrations are commonly employed to mitigate the turbulent boundary layer trailing-edge noise. However, significant discrepancies persist between model predictions and experimental observations. In this paper, we show that this results from the frozen turbulence assumption. A fully-developed turbulent boundary layer over a flat plate is first simulated using the large eddy simulation (LES) method, with the turbulence at the inlet generated using the digital filter method (DFM). The space-time correlations and spectral characteristics of wall pressure fluctuations are examined. The simulation results demonstrate that the coherence function decays in the streamwise direction, deviating from the constant value of unity assumed in the frozen turbulence assumption. By considering an exponential decay function, we relax the frozen turbulence assumption and develop a prediction model that accounts for the intrinsic non-frozen nature of turbulent boundary layers. To facilitate a direct comparison with frozen models, a correction coefficient is introduced to account for the influence of non-frozen turbulence. The comparison between the new and original models demonstrates that the new model predicts lower noise reductions, aligning more closely with the experimental observations. The physical mechanism underlying the overprediction of the noise model assuming frozen turbulence is discussed. The overprediction is due to the decoherence of the phase variation along the serrated trailing edge. Consequently, the ratio of the serration amplitude to the streamwise frequency-dependent correlation length is identified as a crucial parameter in determining the correct prediction of far-field noise.

\end{abstract}


\section{Introduction}

 Trailing-edge (TE) noise is a major concern in various industrial applications, including wind turbines, cooling fans, and turbo-machinery. As a turbulent boundary layer convects past a trailing edge, pressure fluctuations beneath the boundary layer scatter into sound, leading to noise emissions. Inspired by the silent flight of owls \citep{jaworski2020aeroacoustics}, serrated trailing edges have emerged as a promising approach to reduce this noise.
 
 Extensive research has been conducted to investigate the effectiveness of serrated trailing edges in reducing noise. \citet{howe1991aerodynamic,howe1991noise} first developed an analytical model to predict the scattered noise from serrated trailing edges and found that sharp sawtooth serrations are effective in suppressing TE noise. However, the experimental studies conducted by \citet{gruber2012airfoil} showed that Howe's model significantly overpredicted the noise reduction due to TE serrations. Based on a Fourier expansion technique, \citet{lyu2016prediction} extended Amiet's theory \citep{amiet1976noise} to sawtooth trailing edges. Lyu's model yielded a more realistic noise reduction prediction compared to Howe's model. It was shown that the mechanism underlying the noise reduction was the destructive interference effects of the scattered pressure due to the presence of serrations. Two parameters were identified to be important in effectively reducing TE noise. Recently, \citet{ayton2018analytic} presented an analytical model based on the Wiener-Hopf method and applied this model to five test-case TE geometries. Furthermore, \citet{lyu2020rapid} simplified Ayton's model by approximating the infinite interval involving two infinite sums, and the resulting model consumes much less time when evaluated. However, as a semi-infinite flat plate was assumed, the solution was strictly two-dimensional (2D) (the predicted sound pressure decays as $1/\sqrt{r}$ where $r$ is the cylindrical radial distance of the observer), rendering it inapplicable to applications involving rotating blades \citep{lyu2023analytical}. In most practical applications, however, blades are in a state of rotation during their operation, such as the propellers of drones. \myadd{\citet{halimi2019analytical} investigated the broadband noise from a small remotely piloted aircrafts (RPAs) propeller with
 sawtooth serrations using the first-order approximation of Lyu's model.} \citet{tian2022prediction} conducted a theoretical investigation on the noise emitted from three kinds of rotating serrated blades using the second-order approximation. The three-dimensional (3D) directivity patterns of an isolated flat plate were found to be important for the far-field noise characteristics of a rotating blade. 
 
 Despite the valuable insights provided by theoretical models, indispensable discrepancies still exist between the latest analytical predictions and experimental results \citep{arce2016flow,zhou2020study, oerlemans2009reduction}. In a recent study, \citet{zhou2020study} conducted anechoic wind tunnel experiments to investigate the effect of serration shape and flexibility on TE noise. Their findings demonstrated that the new analytical model proposed by \citet{lyu2020rapid} still notably overpredicted the noise reduction capacity achieved by serrations. Given that all these analytical prediction models rely on the statistics of the wall pressure fluctuations as inputs, an accurate characterization of these fluctuations is crucial to an accurate noise prediction.
 
 Comprehensive reviews on the features of the wall pressure fluctuations can be found in the works of \citet{willmarth1975pressure} and \citet{bull1996wall}. In general, the temporal and spatial characteristics of the wall pressure fluctuations on a flat plate can be described in terms of the wavenumber-frequency spectrum, which exhibits two distinct regions. The first region is called the acoustic domain \citep{gloerfelt2013turbulent,blake2012mechanics} where the phase speed is equal to or greater than the speed of sound, enabling efficient radiation to the far field. The second region, referred to as the convected domain, comprises wave components that travel at speeds slower than the speed of sound. Pressure fluctuations in the convected domain exhibit significantly higher magnitudes compared to those observed in the acoustic domain and are related to the scattering process of the TE noise. In practice, semi-empirical wall pressure spectrum models are commonly used in analytical noise predictions, such as the Corcos model \citep{corcos1964structure}, the Chase model \citep{chase1987character}, and the Goody model \citep{goody2004empirical}. Semi-empirical models are usually empirically formulated according to certain scaling laws. \citet{hwang2009comparison} conducted a comparison of the frequency spectra calculated using nine semi-empirical models and found that the Goody model could provide the best overall estimation. Recently, the TNO model has shown promise in obtaining the surface pressure wavenumber-frequency spectrum \citep{stalnov2016towards} and could potentially improve noise prediction performance compared to other empirical models \citep{mayer2019semi}. 
 
 One of the most important assumptions made in modelling the turbulent flow is Taylor's hypothesis of frozen turbulence \citep{taylor1938spectrum}. Taylor hypothesized that the spatial patterns of turbulent motions are carried past a fixed point at the convection velocity without changing significantly. The frozen turbulence assumption depicted a simple scenario that could provide significant convenience in developing analytical models. However, the applicability of this assumption was open to some debate. \citet{lin1953taylor} has shown that this hypothesis is not applicable in cases of high shear flows, such as turbulent boundary layers and the mixing region of a jet. The large-scale shear flows induce the distortions of small eddies as they are carried downstream \citep{zhao2009space}. Subsequently, numerous studies have focused on assessing the validity and improving Taylor's hypothesis \citep{wills1964convection,fisher1964correlation,dennis2008limitations,del2009estimation,renard2015scale,he2017space}. \citet{fisher1964correlation} pointed out that when intensity is high, different turbulent spectral components appear to travel at different speeds. \myadd{Furthermore, under the frozen turbulence assumption, the energy spectrum obtained in a frame of reference moving with the convection velocity contains only components of zero frequency. However, the experimental results of \citet{fisher1964correlation} showed that the energy was spread over a considerable band of frequencies for the shear flows.} In the region of high shear stress within a turbulent channel flow, \citet{del2009estimation} showed that the  phase velocity of the modes with long wavelengths was higher than the local mean velocity. They also proposed a method to determine the convection velocity that relies solely on the spectral information in the temporal or spatial direction. Considering the frozen turbulence assumption implied a first-order approximation, \citet{he2006elliptic} developed an elliptic model based on a second-order approximation. Two characteristic velocities were utilized in this model, i.e., a convection velocity and a velocity that characterizes the distortion of flow patterns. The elliptic model can be used to reconstruct space-time correlations from temporal correlations and has been validated in turbulent channel flows \citep{zhao2009space}, turbulent boundary layers \citep{wang2014trpiv}, and turbulent Rayleigh-B\'enard convection (RBC) \citep{he2012transition}. 

 For the prediction of TE noise, most previous analytical models adopted the frozen turbulence assumption to facilitate a quick estimation, which may be a potential contributor to the discrepancies between models and experiments. Recently, several experimental and numerical studies \citep{avallone2016three,avallone2018noise,zhou2020study,pereira2022physics} have called into question the use of the frozen turbulence assumption. Therefore, it is necessary to explore to which extent the frozen turbulence assumption approximates the real turbulent statistics and to develop methods for incorporating the non-frozen effect in noise prediction models.
 
 This paper is structured as follows. Section \ref{ss2} shows a statistical description of wall pressure fluctuations. Section \ref{ss3} describes the numerical setup employed to simulate a fully-developed turbulent boundary layer. The correlation and spectral features of wall pressure fluctuations are examined. Subsequently, in $\S$\ref{ss4}, a new model that accounts for the non-frozen effect is proposed, and the corresponding prediction results are presented. Section \ref{ss5} elucidates the physical mechanism behind the noise reduction when non-frozen turbulence is taken into consideration. The final section concludes the present paper and lists our future work.
 
\section{The statistical description of the wall pressure fluctuations}\label{ss2}

In this paper, we shall consider a turbulent boundary layer that develops on a flat plate under a zero mean pressure gradient. In TE noise modelling, the statistical spectrum of the wall pressure fluctuations beneath a turbulent boundary layer is often used as an input. We define some of the key quantities in this section. The space-time correlation of the wall pressure fluctuations $p^\prime(\boldsymbol{x},t)$ at position $\boldsymbol{x}=(x,z)$ at time $t$ is defined by
\begin{equation}
	Q_{pp}(\boldsymbol{x},t;\boldsymbol{\xi},\tau) = \langle p^\prime(\boldsymbol{x},t)p^\prime(\boldsymbol{x}+\boldsymbol{\xi},t+\tau)\rangle,
\end{equation}
where $\boldsymbol{\xi}$=$(\xi,\eta)$, $\xi$ and $\eta$ are the spatial separations in the streamwise and spanwise directions respectively, and $\tau$ is the time delay. As the turbulent boundary layer develops slowly in the streamwise direction, the flow field may be regarded as homogeneous in the directions parallel to the wall and stationary in time within the scales of interest. Thus, we have $Q_{pp}(\boldsymbol{x},t;\boldsymbol{\xi},\tau)$$\approx$$Q_{pp}(\boldsymbol{\xi},\tau)$. The correlation coefficient is then defined by
\begin{equation}
	R_{pp}(\xi,\eta,\tau) = \dfrac{Q_{pp}(\xi,\eta,\tau)}{Q_{pp}(0,0,0)}.
\end{equation}
The streamwise and spanwise integral lengths can be defined as
\begin{equation}
\Lambda_x = \int_{0}^{\infty}R_{pp}(\xi,0,0)\mathrm{d}\xi,
\label{Lambda1}
\end{equation}
\begin{equation}
\Lambda_z = \int_{0}^{\infty}R_{pp}(0,\eta,0)\mathrm{d}\eta.
\label{Lambda2}
\end{equation}

The spectral density of wall pressure fluctuations can be obtained by performing the Fourier transform of the space-time correlation. In the frequency domain, the single-point spectrum $\phi(\omega)$ is expressed as
\begin{equation}
	\phi(\omega)=\dfrac{1}{2\mathrm{\gpi}}\int_{-\infty}^{\infty}Q_{pp}(0,0,\tau)\mathrm{e}^{-\mathrm{i}\omega \tau}\mathrm{d}\tau,
\end{equation}
where $\omega=2\gpi f$ is the angular frequency and $f$ is the frequency. Similarly, the cross-spectral density is defined by
\begin{equation}
	G_{pp}(\xi,\eta,\omega)=\dfrac{1}{2\mathrm{\gpi}}\int_{-\infty}^{\infty}Q_{pp}(\xi,\eta,\tau)\mathrm{e}^{-\mathrm{i}\omega\tau}\mathrm{d}\tau.
	\label{CrosSpec}
\end{equation}

Making use of the single-point spectrum and the cross-spectral density, the coherence function can be defined as
\begin{equation}
	\gamma^2(\xi,\eta,\omega)=\dfrac{|G_{pp}(\xi,\eta,\omega)|^2}{\phi(\omega)^2}.
	\label{cohefunc}
\end{equation}
In (\ref{Lambda1}) and (\ref{Lambda2}), we have defined the streamwise and spanwise integral lengths based on the correlation coefficients, which are independent of frequency. However, it is known that the spatial correlation of the pressure fluctuations varies with frequency. Therefore, we introduce the frequency-dependent correlation lengths defined as
\begin{equation}
	l_x(\omega)= \int_{0}^{\infty}\gamma(\xi,0,\omega)\mathrm{d}\xi,
	\label{lx}
\end{equation}
\begin{equation}
	l_z(\omega)= \int_{0}^{\infty}\gamma(0,\eta,\omega)\mathrm{d}\eta.
	\label{lz}
\end{equation}
As will be seen, the characteristics of $l_{x,z}$ differ significantly from those of $\Lambda_{x,z}$ and they have significant implications in noise predictions.

To obtain the wavenumber-frequency spectrum of the wall pressure fluctuations, we perform spatial Fourier transforms on the cross-spectral density, resulting in the definition of the wavenumber-frequency spectrum,
\begin{equation}
	\Pi(k_1,k_2,\omega)=\dfrac{1}{(2\gpi)^2}\int_{-\infty}^{\infty}\int_{-\infty}^{\infty}G_{pp}(\xi,\eta,\omega)\mathrm{e}^{\mathrm{i}(k_1\xi+k_2\eta)}\mathrm{d}\xi\mathrm{d}\eta,
\end{equation}
where $k_1$ and $k_2$ are wavenumbers in the streamwise and spanwise directions, respectively.

The wavenumber-frequency spectrum describes the spectral distribution of energy in wall pressure fluctuations and serves as a key input in TE noise prediction models. The highest levels of pressure fluctuations typically occur within a specific region centered around $k_1=\omega/U_c$, $k_2=0$, where $U_c$ is the convection velocity. This region is the so-called convective ridge. The idea that slowly distorting eddies are convected downstream by the mean flow at a fixed velocity is useful in the study of turbulent shear flows, and is particularly important in the research of aerodynamic noise \citep{wills1964convection}. Various approaches exist for defining the convection velocity \citep{hussain1981measurements}, and a comprehensive review of the convection velocity datasets in turbulent shear flows was conducted by \citet{renard2015scale}. In general, the convection velocity should not be treated as a constant value. This is because eddies of different sizes can convect at different velocities, and so do the eddies with different time scales. Therefore, in general, the convection velocity can be expressed as a function of time delay $\tau$ and streamwise separation $\xi$, or as a function of frequency $\omega$ and streamwise wavenumber $k_1$ in the spectral domain.


Obtaining a well-resolved space-time flow field database is often challenging or impractical in real-world scenarios. As a result, it becomes necessary to reconstruct the wavenumber-frequency spectrum from either the space or time datasets based on the statistical characteristics of the turbulent boundary layer. In 1938, \citet{taylor1938spectrum} proposed the well-known hypothesis that turbulent eddies convect uniformly and unchangingly past a fixed point as if the spatial patterns of the flow field are ``frozen". Under this frozen turbulence assumption, the correlation function satisfies \citep{bull1967wall}
\begin{equation}
	Q_{pp}(\xi,\eta,\tau)=Q_{pp}(\xi-U_c\tau,\eta,0).
	\label{corrFroz}
\end{equation}

The streamwise coherence function can be readily found as a constant from (\ref{cohefunc}), i.e.
\begin{equation}
	\gamma(\xi,0,\omega)=1.
\end{equation}
 This implies that the eddy patterns exhibit perfect coherence in the streamwise direction at all frequencies as they convect downstream. The frozen turbulence assumption has been widely employed due to its simplicity in modelling the wavenumber-frequency spectrum. However, in a real turbulent boundary layer, the eddies undergo distortions caused by the mean shear \citep{fisher1964correlation}. This implies that the streamwise coherence would decay as $\xi$ increases. In such cases, assuming a constant streamwise coherence function may introduce large errors when used in noise prediction models. Therefore, gaining a more comprehensive understanding of the spatial coherence of wall pressure fluctuations, especially the frequency-dependent correlation lengths, is crucial. In the subsequent section, a numerical investigation will be conducted to examine in detail the characteristics of wall pressure fluctuations. 
 
\section{Numerical simulation of a turbulent boundary layer}\label{ss3}
\subsection{Numerical set-up}

To investigate the space-time correlations and spectral characteristics of wall pressure fluctuations, we use the wall-resolved LES method to simulate a fully-developed turbulent boundary layer over a flat plate. Considering that in many applications where TE noise is important the Mach number is relatively low, we choose to perform an incompressible simulation. Compared to the direct numerical simulation (DNS) method, LES requires less number of grid points for wall-resolved simulations \citep{schlatter2010simulations}, leading to less computational resource demands. The LES method solves the spatially-filtered Navier-Stokes equations using a subgrid-scale (SGS) model. For incompressible flows, the filtered momentum and continuity equations can be expressed as
\begin{equation}
	\dfrac{\partial \bar{u}_i}{\partial t}+\dfrac{\partial}{\partial x_j}(\bar{u}_i\bar{u}_j)=-\dfrac{1}{\rho_0}\dfrac{\partial\bar{p}}{\partial x_i}+\nu\dfrac{\partial^2\bar{u}_i}{\partial x_j\partial x_j}-\dfrac{\partial \tau_{ij}}{\partial x_j}, i=1,2,3,
\end{equation}
\begin{equation}
	\dfrac{\partial \bar{u}_j}{\partial x_j}=0, 
\end{equation}
where the overbar denotes filtered variables with a filter width $\Delta$, $t$ is the time, $u_i$ is the velocity component in the $x_i$-direction (also denoted as $u$, $v$, or $w$), $\rho_0$ is the density, $p$ is the pressure, and $\nu$ is the kinematic viscosity. The contributions from sub-grid scales are represented through the SGS stresses $\tau_{ij}=\overline{u_iu_j}-\bar{u}_i\bar{u}_j$, which need to be modeled. Following the eddy-viscosity assumption, the SGS stress can be modelled as
\begin{equation}
	\tau_{ij}-\dfrac{1}{3}\delta_{ij}\tau_{kk}=-2\nu_T\bar{S}_{ij},
\end{equation}
where $\delta_{ij}$ is Kronecker's delta, and $\bar{S}_{ij} = (\partial \bar{u}_i/\partial x_j+\partial \bar{u}_j/\partial x_i)/2$ is the large-scale strain-rate tensor. The SGS viscosity $\nu_{T}$ can be calculated using various models. In this study, the wall-adapting local eddy-viscosity model (WALE) \citep{nicoud1999subgrid} is used due to its ability to account for the wall effect on the turbulent structure. The value of $\nu_T$ is obtained as
\begin{equation}
	\nu_T =C_w^2\Delta^2\dfrac{(S_{ij}^dS_{ij}^d)^{\frac{3}{2}}}{(\bar{S}_{ij}\bar{S}_{ij})^{\frac{5}{2}}+(S_{ij}^dS_{ij}^d)^{\frac{5}{4}}},
\end{equation}
where $C_w$ is a constant coefficient and $S_{ij}^d$ is the traceless symmetric part of the square of the velocity gradient tensor,
\begin{equation}
	S_{ij}^d=\dfrac{1}{2}\left(\bar{g}_{ij}^2+\bar{g}_{ji}^2\right)-\dfrac{1}{3}\delta_{ij}\bar{g}_{kk}^2.
\end{equation}
Here, $\bar{g}_{ij}={\partial \bar{u}_i}/{\partial x_j}$ and $\bar{g}_{ij}^2=\bar{g}_{ik}\bar{g}_{kj}$.

For the turbulent inlet boundary condition, we use the synthetic turbulent inflow generator. The generator employs a 2D filter to produce spatially correlated 2D slices of data. The instantaneous velocity on the slice is computed as
\begin{equation}
	u_i=\bar{u}_i+a_{ij}\Psi_j,
\end{equation}
where $\Psi_j$ denotes the filtered fluctuating velocity field and $a_{ij}$ is the amplitude tensor, which is related to the Reynolds stresses tensor $R_{ij}$ by 
\begin{equation}
	a_{ij} = 
	\begin{bmatrix}
		\sqrt{R_{11}}& 0 & 0\\
		\dfrac{R_{21}}{a_{11}}& \sqrt{R_{22}-a_{21}^2} & 0\\
		\dfrac{R_{31}}{a_{11}}& \dfrac{R_{32}-a_{22}a_{31}}{a_{22}} & \sqrt{R_{33}-a_{31}^2-a_{32}^2}\\
	\end{bmatrix}.
\end{equation}

By applying spatial and temporal filters to the random array sequences, we can introduce the desired temporal and spatial correlations in the instantaneous velocity fluctuations. The spatial turbulent length scale is defined by the two-point correlation, which is given by
\begin{equation}
	L_i^j(\boldsymbol{x})=\int_0^\infty\dfrac{\overline{u_i'(\boldsymbol{x})u_i'(\boldsymbol{x}+\boldsymbol{e}_jr)}}{\overline{u_i'(\boldsymbol{x})u_i'(\boldsymbol{x})}}\mathrm{d}r,
\end{equation}
where $r$ is the spatial separation in the $j$-direction and $u_i'(\boldsymbol{x})$ denotes the velocity fluctuations. The parameters required to generate a turbulent inlet condition using DFM include the profiles of the mean velocity, turbulent Reynolds stresses, and turbulent length scales. These parameters can be obtained through various approaches, such as a precursor DNS simulation, modelling from a Reynolds-averaged Navier-Stokes (RANS) computation, or measurements from experiments. In this work, the mean velocity and turbulent Reynolds stresses are obtained from the DNS data provided by \citet{schlatter2010assessment}. The Reynolds number at the inlet, based on the momentum thickness $\theta$ and the free-stream velocity $U_0$, is set to 1410. Regarding the turbulent length scale $L_u^x$, a constant value may be prescribed. \myadd{Following \citet{wang2022influence}, $L_u^x$ is set to the boundary layer thickness scaled by a factor of 0.6 and other turbulent length scales can be prescribed based on $L_u^x$.} Table \ref{L} shows the nine turbulent length scales used at the inlet with DFM, where $\delta_{in}$ denotes the boundary layer thickness at the inlet. 

\begin{table}
	\centering
	\begin{tabular}{ccccccccc}
	
		$L_u^x$ & $L_v^x$ & $L_w^x$ & $L_u^y$ & $L_v^y$ & $L_w^y$ & $L_u^z$ & $L_v^z$ & $L_w^z$\\
		0.6$\delta_{in}$ & 0.33 $L_u^x$ & 0.33 $L_u^x$ & 0.235 $L_u^x$ & 0.235 $L_u^x$ & 0.235 $L_u^x$ & 0.35 $L_u^x$ & 0.35 $L_u^x$ & 0.35 $L_u^x$\\
		
	\end{tabular}
	\caption{Turbulent length scales used in the inlet boundary condition.}
	\label{L}
\end{table} 

\begin{figure}
	\centering
	\includegraphics[scale = 0.42]{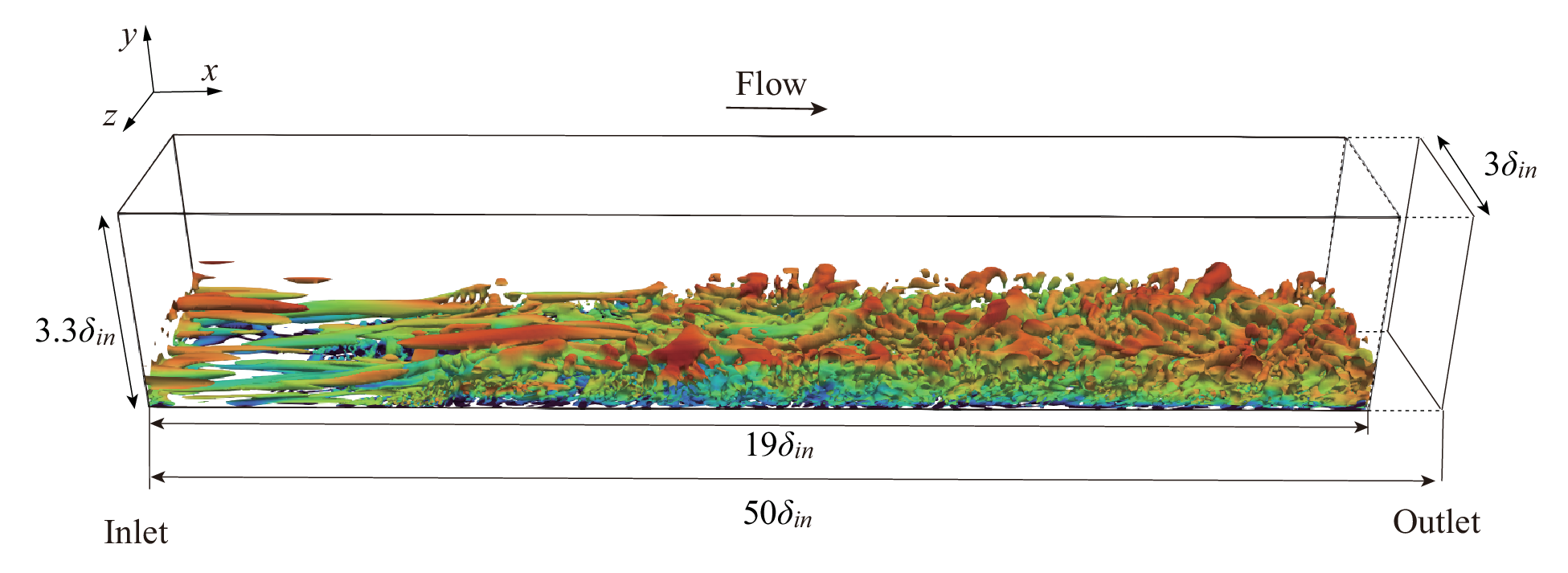}
	\caption{Computational domain of the LES simulation and an instantaneous flow field from the inlet to $19\delta_{in}$ downstream. The flow is visualized using the Q-criterion ($Q=2.5\times10^4$) and colored by the streamwise velocity (with levels increasing, the color changes from blue to red).}
	\label{domain}
\end{figure}

\begin{table}
	\centering
	\begin{tabular}{cccccccccc}
		$Re_{\theta}$ & $L_x/\delta_{in}$ & $L_y/\delta_{in}$ & $L_z/\delta_{in}$ & $n_x$ & $n_y$ & $n_z$ & $\Delta x^+$ & $\Delta y_{min}^+$ & $\Delta z^+$  \\ 
		1410 & $50$ & $3.3$ & $3$ & 2000 & 45 & 120 & 12.3 & 1.1 & 12.3\\
	\end{tabular}
	\caption{Grid information for the LES of a spatially developing turbulent boundary layer.}
	\label{gridinf}
\end{table}

The numerical simulation in this study is conducted using OpenFOAM-v2206. The computational domain, as illustrated in figure \ref{domain}, has dimensions of $50\delta_{in}\times 3.3\delta_{in}\times 3\delta_{in}$ in the streamwise ($x$), wall-normal ($y$), and spanwise ($z$) directions, respectively. The mesh cells are exponentially distributed along the $y$-axis and uniformly placed along the streamwise and spanwise directions. Table \ref{gridinf} lists detailed parameters employed in this study. To ensure grid independence, both fine and coarse meshes are tested, and the results of the grid independence test can be found in Appendix \ref{appa}. 

For the velocity boundary condition, a slip condition is used on the top wall, while a no-slip condition is imposed on the bottom wall. In the lateral direction, a periodic boundary condition is employed to simulate an infinite domain. At the outlet, the inletOutlet boundary condition is imposed. Regarding the pressure boundary conditions, all boundaries are set to zero-gradient except for the top boundary, where a fixed pressure is prescribed.

\subsection{Numerical results}

In this section, we present the simulation results of a spatially-developing turbulent boundary layer using the LES method. The turbulent statistics are obtained after the flow field reaches a statistically-stationary state. The LES data is used to show the spatial evolution of the flow structures, validate the flow statistics, and examine the statistical characteristics of wall pressure fluctuations.


\subsubsection{Flow field}

\begin{figure}
	\centering
	\includegraphics[scale=0.7]{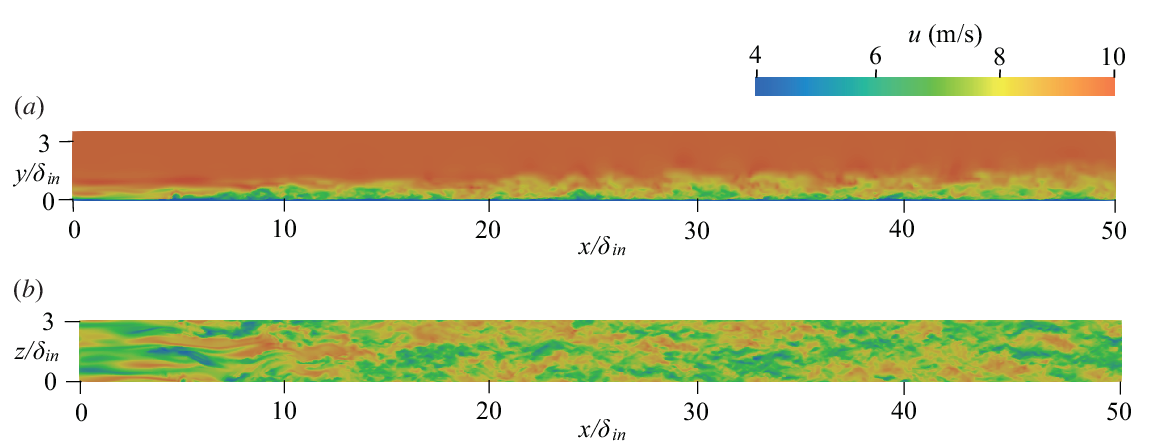}
	\caption{Instantaneous snapshots of the streamwise velocity: (\textit{a}) sideview in the $x$-$y$ plane and (\textit{b}) overview in the $x$-$z$ plane at $y=0.3\delta_{in}$.}
	\label{2-Dflowfield}
\end{figure}

An instantaneous flow field is visualized using an isosurface of the Q-criterion, as shown in figure \ref{domain}. The computational domain is sufficiently long in the streamwise direction, and the flow region from the inlet to $19\delta_{in}$ downstream is selected for visualization. It can be seen that the unsteady flow structures generated at the inlet are not physical. But these unphysical structures quickly decay and the flow becomes more physical after around $10 \delta_{in}$. Similar phenomena can also be seen in figure \ref{2-Dflowfield}, where instantaneous snapshots of the flow field are displayed. Figure \ref{2-Dflowfield}(a) shows the sideview of the instantaneous streamwise velocity field in the $x-y$ plane while figure \ref{2-Dflowfield}(b) shows the overview in the $x-z$ plane located at $y=0.3\delta_{in}$. The visualization demonstrates that artificial turbulent structures are instigated at the inlet, preserved for a short distance downstream, and subsequently replaced by more physical structures.

\begin{figure}
	\centering
	\includegraphics[scale=1]{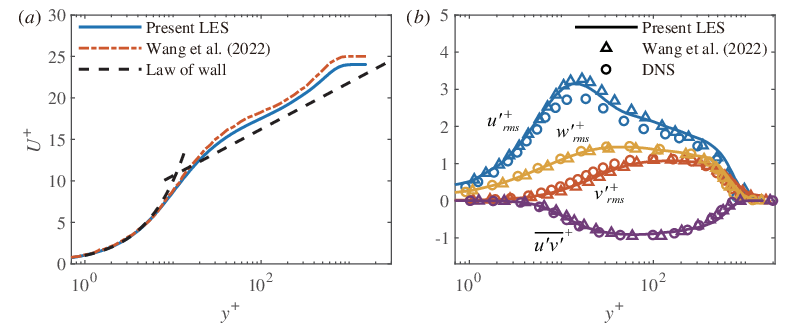}
	\caption{Profiles of mean flow statistics at $x=40\delta_{in}$: (\textit{a}) velocity and (\textit{b}) Reynolds shear stress as well as streamwise, spanwise, and wall-normal velocity fluctuations.}
	\label{Umean}
\end{figure}

Flow statistics are obtained by averaging over the spanwise direction $z$ and time $t$. Therefore, the streamwise velocity can be decomposed into $u=U+u^\prime$, where $U$ and $u^\prime$ denote the mean velocity and the fluctuating velocity, respectively. The wall shear stress can be calculated as $\tau_w=\mu(\mathrm{d}U/\mathrm{d}y)|_{y=0}$, where $\mu$ is the dynamic viscosity. The friction velocity is defined as $u_{\tau}=\sqrt{\tau_w/\rho_0}$, and the characteristic length is given by $l_{\star}=\nu/u_{\tau}$. Therefore, the mean velocity and distance normal to the wall can be expressed in non-dimensional forms as $U^+=U/u_{\tau}$ and $y^+=y/l_{\star}$, respectively. \myadd{In the subsequent analysis, the streamwise location $x=40\delta_{in}$ is used as the reference position}. At this location, the friction velocity is $u_{\tau}=0.041U_0$, and the Reynolds number based on the momentum thickness is $Re_{\theta}=2053$.

Figure \ref{Umean} shows the distribution of the mean velocity and turbulent Reynolds stress components. The simulated mean velocity exhibits good agreement with those obtained by \citet{wang2022influence} using the dynamic Smagorinsky model, as shown in figure \ref{Umean}(a). In the near-wall region, the simulated mean velocity profile collapses well with the linear law. However, in the logarithmic region, both the simulation results of the present study and \citet{wang2022influence} slightly deviate from the log-law. From figure \ref{Umean}(b), we can see that the simulated velocity fluctuations and Reynolds shear stress are in good agreement with DNS results. Slight deviations from the DNS profile can be seen for the ${u^{\prime +}_{rms}}$ profile in both the present study and the work of \citet{wang2022influence}, which might be attributed to the artificial inflow-boundary condition employed in these two works. Nevertheless, figure \ref{Umean} shows the present LES captures essential flow physics.

\subsubsection{Properties of wall pressure fluctuations}

In this section, we examine the correlations, spectral properties, and convection characteristics of wall pressure fluctuations.

\begin{figure}
	\centering
	\hspace*{0.2cm}\includegraphics[scale=1]{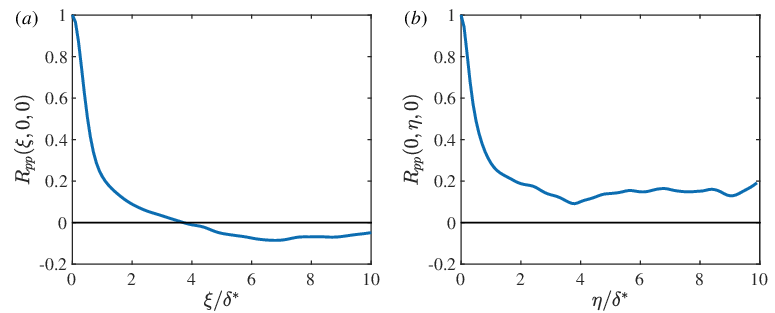}
	\caption{One-dimensional spatial correlations: (\textit{a}) streamwise direction and (\textit{b}) spanwise direction.}
	\label{Onecorr}
\end{figure}

\begin{figure}
	\centering
	\includegraphics[scale=1]{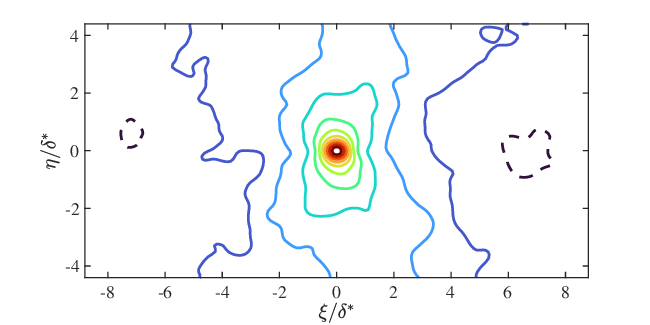}
	\caption{Contours of the spatial correlation $R_{pp}(\xi,\eta,0)$. Solid lines denote positive isocontours: 0, 0.1, 0.2, 0.3, 0.4, 0.5, 0.6, 0.7, 0.8, and 0.9; and the dashed line denotes a negative isocontour of -0.1.}
	\label{Twocorr}
\end{figure}

Figure \ref{Onecorr} shows the spatial correlations in the streamwise and spanwise directions, respectively. It can be seen that both correlations decay rapidly with increasing separation. However, the spanwise correlation remains positive throughout the shown range, while the streamwise correlation changes sign at $\xi/\delta^\ast\approx3.9$, which aligns with the findings of \citet{bull1967wall}. The decay of the correlations may be improved by using a larger computational domain but should suffice for the present study. Figure \ref{Twocorr} presents the contour plot of the two-point spatial correlation. The overall pattern is similar to the observations reported by \citet{bull1967wall}, where the contours are nearly circular for small separations, indicating near isotropy of the field. However, as the separations increase, the contours elongate in the spanwise direction, indicating increasing anisotropy. This elongation is likely attributed to large-scale flow structures and implies that $\Lambda_z$ is larger than $\Lambda_x$. In the streamwise direction, negative areas can be seen, consistent with the behavior depicted in figure \ref{Onecorr}(a).

\begin{figure}
	\centering
	\includegraphics[scale=1]{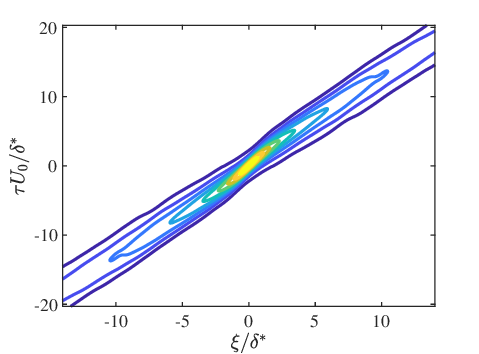}
	\caption{Contour of the space-time correlation $R_{pp}(\xi,0,\tau)$, levels are from 0.1 to 0.9 with an increment of 0.1.}
	\label{spatim}
\end{figure}

\begin{figure}
	\centering
	\includegraphics[scale=1]{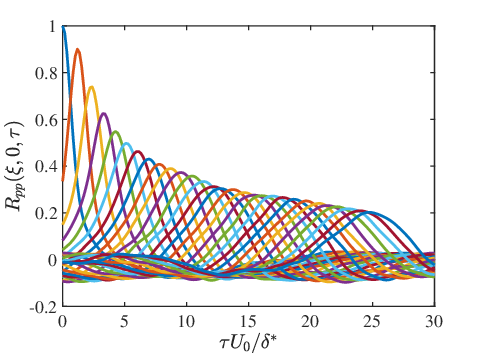}
	\caption{Space-time correlation as a function of time delay for various fixed streamwise separations with an increment of $0.7\delta^\ast$.}
	\label{figurespax}
\end{figure}

An important characteristic of non-frozen turbulence is the presence of elliptic patterns in the contours of the space-time correlation $R_{pp}(\xi,0,\tau)$, as illustrated in figure \ref{spatim}. In contrast, under the assumption of frozen turbulence, the contours degrade to parallel straight lines. As shown in figure \ref{spatim}, the concentration of contour lines into a narrow band suggests that the development of flow structures downstream includes both convection and decay. Figure \ref{figurespax} presents the space-time correlations for various fixed streamwise separations as a function of time delay. It can be seen that the correlation peak decreases as the streamwise separation increases. This behavior indicates a decaying correlation between the pressure fluctuations as separation distance increases. This contrasts directly with a non-decaying correlation implied in the frozen turbulence assumption (see (\ref{corrFroz})).

\begin{figure}
	\centering
	\includegraphics[scale=1]{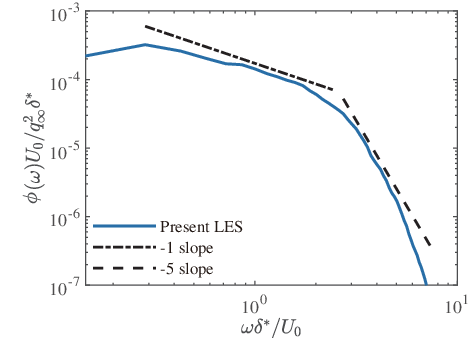}
	\caption{Power spectra density scaled with outer parameters.}
	\label{PSD}
\end{figure}

Figure \ref{PSD} shows the power spectral density (PSD) of the pressure fluctuations obtained using Welch's method \citep{welch1967use} and normalized by the dynamic pressure $q_{\infty}=\rho_0U_0^2/2$ and the displacement thickness $\delta^\ast$. The simulated pressure spectrum exhibits two distinct regimes: a -1 scaling regime and a -5 scaling regime. The -1 scaling is associated with the eddies present in the logarithmic region of the boundary layer. These eddies contribute to the energy distribution in the low-frequency range of the spectrum. On the other hand, the -5 scaling, appeared in the high-frequency range, is related to the presence of smaller-scale eddies within the buffer layer \citep{blake2012mechanics}.

\begin{figure}
		\centering
		\includegraphics[scale=1]{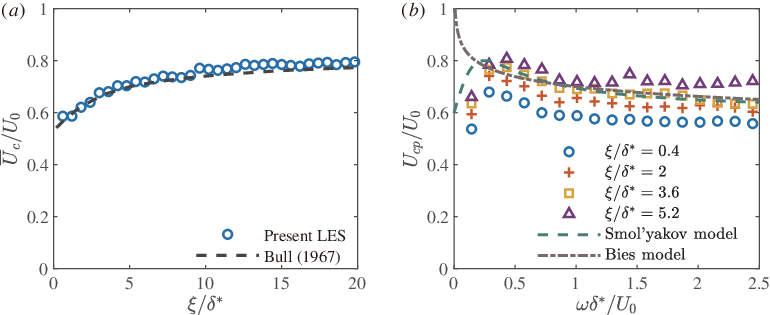}
		\caption{(\textit{a}) Comparison of the mean convection velocity. (\textit{b}) Phase velocity as a function of frequency for fixed streamwise separations with an increment of $1.6\delta^\ast$ as well as convection velocities calculated using the Smol'yakov model and the Bies model.}
		\label{convel}
\end{figure}

The mean convection velocity $\overline{U}_c$ can be estimated by determining the time delay $\tau$ corresponding to the correlation peak shown in figure \ref{figurespax} for a fixed spatial separation $\xi$, i.e.
	\begin{equation}
		\overline{U}_c(\xi)=\dfrac{\xi}{\tau}.
	\end{equation}
On the other hand, to obtain a frequency-dependent convection velocity, we can use the cross-spectral density $G_{pp}(\xi,\eta,\omega)$, which is a complex function \citep{gloerfelt2013turbulent}. Let $\theta_p(\xi,\omega)$ denotes the phase of $G_{pp}(\xi,0,\omega)$, then the phase velocity in the streamwise direction $U_{cp}(\xi,\omega)$ can be determined by \citep{farabee1991spectral}
	\begin{equation}
		U_{cp}(\xi,\omega)=-\dfrac{\omega \xi}{\theta_p(\xi,\omega)}.
	\end{equation}
The cross-spectral density can be written as
\begin{equation}
	G_{pp}(\xi,\eta,\omega)=|G_{pp}(\xi,\eta,\omega)|\mathrm{e}^{-\mathrm{i}\xi\omega/U_{cp}(\xi,\omega)}.
	\label{GppCom}
\end{equation}
The exponential term in (\ref{GppCom}) represents the convection behavior of turbulent eddies, where the convection velocity is expected to be dependent on the frequency $\omega$ and the separation distance $\xi$.

Figure \ref{convel}(a) shows the comparison between the simulated mean convection velocity and the experimental measurements by \citet{bull1967wall}. It can be seen that there is good agreement between the two results. As the streamwise separation increases, the mean convection velocity also increases and approaches 0.8. Figure \ref{convel}(b) shows the variations of the phase velocity as a function of frequency for various fixed streamwise separations. The convection velocities calculated using the empirical models proposed by \citet{smol2006new} and \citet{bies1966review} (see Appendix \ref{appb}) are also shown for comparison. It is evident that the phase velocity increases with increasing streamwise separation, and as the frequency increases, the phase velocity rises rapidly, reaches a peak velocity, and then decays slowly. This suggests that the assumption of frozen turbulence, which assumes that all eddies in the turbulent boundary layer convect at the same velocity, is not strictly valid \citep{farabee1991spectral}. Both the Smol'yakov model and the Bies model agree with the numerical results at intermediate and high frequencies, but the Bies model fails to capture the characteristics at low frequencies. 

\begin{figure}
	\centering
	\includegraphics[scale=0.95]{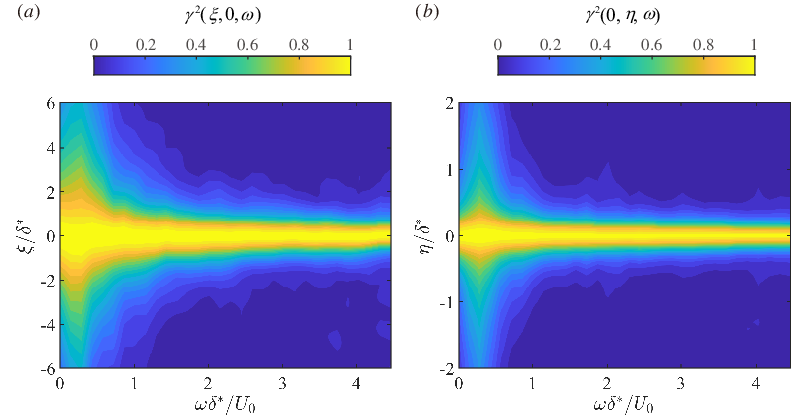}
	\caption{Streamwise (\textit{a}) and spanwise (\textit{b}) coherences of wall pressure fluctuations.}
	\label{strcoh}
\end{figure}

Figure \ref{strcoh} examines the streamwise and spanwise coherences of pressure fluctuations. We see that the contour shapes of the coherence functions in the two directions are similar. For a fixed non-zero separation, the coherence increases with the increase of frequency and then decays. This behavior indicates that the low-frequency components, associated with large-scale structures, maintain their coherence over longer distances, while the high-frequency components lose their coherence more rapidly with increasing separation. Particularly, for the non-dimensional frequency $\omega\delta^\ast/U_0>1$, the wall pressure fluctuations quickly lose their coherence as the separation increases, indicating that the perfect coherence assumed by the frozen turbulence might lead to significant errors. The coherence contours provide valuable information for determining the frequency-dependent correlation lengths.  

\begin{figure}
	\centering
	\includegraphics[scale=1]{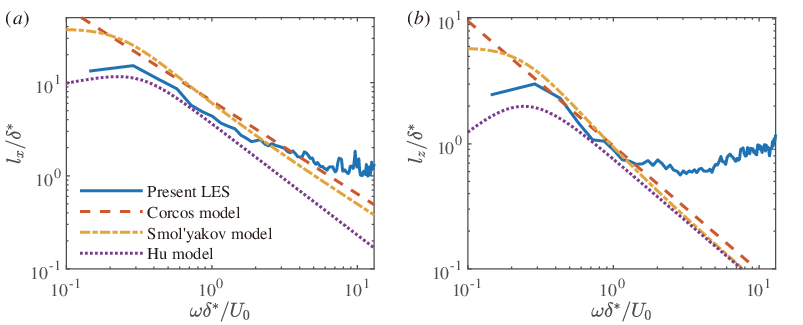}
	\caption{Frequency-dependent correlation lengths : (\textit{a}) streamwise direction and (\textit{b}) spanwise direction.}
	\label{cohlen}
\end{figure}

Equations (\ref{lx}) and (\ref{lz}) provide the definitions of the streamwise and spanwise frequency-dependent correlation lengths. However, in practical applications, curve-fitting approaches are often employed. For each discrete frequency, the frequency-dependent correlation lengths can be assumed in exponential forms, i.e. 
\begin{equation}
	\gamma(\xi,0,\omega)=\mathrm{e}^{-|\xi|/l_x(\omega)},
	\label{gammax}
\end{equation}
\begin{equation}
	\gamma(0,\eta,\omega)=\mathrm{e}^{-|\eta|/l_z(\omega)}.
	\label{gammaz}
\end{equation}   

In figure \ref{cohlen}, the frequency-dependent correlation lengths in the streamwise and spanwise directions are plotted as functions of frequency. Three empirical models, i.e. the Corcos model \citep{corcos1964structure}, the Smol'yakov model \citep{smol2006new}, and the Hu model \citep{hu2021coherence} are also shown for comparisons, whose formulations can be found in Appendix \ref{appc}. It can be seen that both correlation lengths increase slightly as frequency increases in the low-frequency range. When the frequency further increases, $l_x(\omega)$ and $l_z(\omega)$ decay rapidly. All three empirical models exhibit similar decay trends within the intermediate- and high-frequency ranges. However, the Smol'yakov model and the Hu model could capture the characteristics at low frequencies. In addition, at higher frequencies, we can see that the simulated correlation lengths decay slowly and even increase. This phenomenon can also be found in the study of \citet{van2015estimation} and a mesh refinement may be helpful to obtain improved decay tendencies of the frequency-dependent correlation lengths in this regime.

An interesting observation is that for the same frequency, $l_x(\omega)>l_z(\omega)$. This is in contrast to the streamwise frequency-independent correlation length $\Lambda_x$, which is smaller than the spanwise correlation length $\Lambda_z$. This phenomenon can be attributed to the convection of turbulence in the streamwise direction. Considering the definition of the frequency-dependent correlation length, we have
\begin{equation}
 l_x(\omega)=\int_{0}^{\infty}\dfrac{|G_{pp}(\xi,0,\omega)|}{\phi(\omega)}\mathrm{d}\xi.
 \label{lx2}
\end{equation}

From (\ref{lx2}), we can see that $l_x(\omega)$ characterizes the correlation that eliminates the effect of the streamwise convection of turbulent eddies. Physically, this represents the correlation length measured in a coordinate frame that moves with the eddy.

Introducing the complex form of the cross-spectral density, the space-time correlation in the streamwise direction can be written as
\begin{equation}
	Q_{pp}(\xi,0,\tau)=\int_{-\infty}^{\infty}|G_{pp}(\xi,0,\omega)|\mathrm{e}^{\mathrm{i\omega}(\tau-\xi/U_{cp}(\xi,\omega))}\mathrm{d}\omega.
\end{equation}
Therefore, the frequency-independent correlation length reads
\begin{equation}
	\Lambda_x=\dfrac{1}{Q_{pp}(0,0,0)}\int_{0}^{\infty}\int_{-\infty}^{\infty}|G_{pp}(\xi,0,\omega)|\mathrm{e}^{\mathrm{i\omega}(\tau- \xi/U_{cp}(\xi,\omega))}\mathrm{d}\omega\mathrm{d}{\xi}.
	\label{Lambda3}
\end{equation}

Comparing (\ref{lx2}) and (\ref{Lambda3}), it is clear that the calculation of the streamwise frequency-independent correlation length $\Lambda_x$ takes into account the influence of the convection of turbulent structures. On the other hand, the frequency-dependent correlation length $l_x(\omega)$ does not. Since the convection of eddies contributes to the decay of the correlation, it is possible that $\Lambda_x < \Lambda_z$ even though $l_x(\omega)>l_y(\omega)$. 

In this part, we have examined two important features of wall pressure fluctuations that are omitted by the frozen turbulence assumption. First, the convection velocity is not strictly constant. Second, the eddies lose their coherence as they convect downstream, leading to a finite streamwise correlation length. These two characteristics are important non-frozen properties and must be accounted for in the modelling of the far-field noise emitted from serrated trailing edges.

\section{Acoustic prediction}\label{ss4}
	
\subsection{Model establishment}\label{ss41}

\begin{figure}
	\centering
	\includegraphics[scale=1]{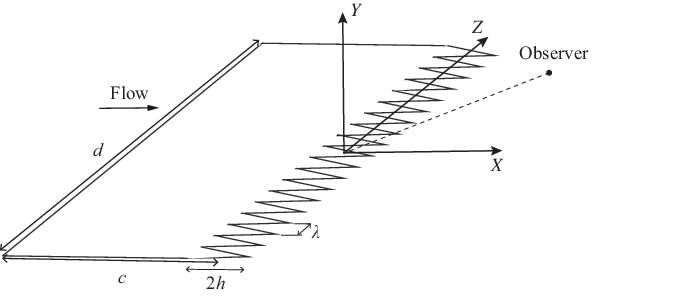}
	\caption{The schematic of a flat plate with TE serrations.}
	\label{flatp}
\end{figure}

With the properties of wall pressure fluctuations and the numerical results discussed above, we are in a position to consider the influence of non-frozen turbulence on the noise prediction for serrated trailing edges. As shown in figure \ref{flatp}, consider a flat plate encountering a uniform flow. The plate has a chord length of $c$, a span of $d$, and a trailing edge with serrations of amplitude $2h$ and wavelength $\lambda$.

According to \citet{lyu2016prediction}, for a general wall pressure fluctuation characterized by its wavenumber-frequency spectrum $\Pi(k_1, k_2, \omega)$, the far-field acoustic power spectral density $S_{pp}$ at the observer position $\boldsymbol{X}_0=(X_0,Y_0,Z_0)$ is found to be 
 \begin{equation}
	S_{pp}(\boldsymbol{X}_0,\omega)=\left(\dfrac{\omega Y_0c}{4\gpi c_0S_0^2}\right)^22\gpi d\sum_{m=-\infty}^{\infty}\int_{-\infty}^{\infty}|\mathcal{L}(k_1,2m\gpi/\lambda,\omega)|^2\Pi(k_1,2m\gpi/\lambda,\omega)\mathrm{d}k_1,
	\label{predicfull}
\end{equation}
where $c_0$ is the speed of sound, $S_0^2=X_0^2+(1-M_0^2)(Y_0^2+Z_0^2)$, and $M_0$ is the Mach number of the flow. $\mathcal{L}$ is the response function which could be calculated iteratively. Note no assumption regarding the frozen property of the wall pressure fluctuations is assumed here. More details about this model can be found in \citet{lyu2016prediction}. 

It can be seen from (\ref{predicfull}) that the formulation relies on the wavenumber-frequency spectrum as input and involves an infinite integral over the streamwise wavenumber $k_1$. While this integral can be evaluated numerically, such an approach would be computationally demanding. Furthermore, obtaining an accurate wavenumber-frequency spectrum $\Pi(k_1,k_2,\omega)$ is challenging in both numerical simulations and experimental measurements. To achieve a convenient prediction and determine the physical impact of non-frozen turbulence on TE noise, we aim to develop a simplified model based on the characteristics of wall pressure fluctuations. Specifically, we approximate the cross-spectral density using a variable-separation form,
\begin{equation}
	G_{pp}(\xi,\eta,\omega)=\gamma(\xi,0,\omega)\mathrm{e}^{-\mathrm{i}\xi\omega/U_c(\omega)}G_{pp}(0,\eta,\omega).
	\label{sepa}
\end{equation}

This form is similar to the Corcos model \citep{corcos1964structure}, which was initially developed by fitting experimental data and has since been widely used in the modelling of wall pressure fluctuations. However, the Corcos model is more stringent as it assumes that the normalized cross-spectral density can be represented by a function that depends on a single dimensionless variable. Here, the convection velocity is expressed as a function of the angular frequency, while its dependency on the streamwise separation is neglected for simplicity. 

By performing Fourier transforms, the wavenumber-frequency spectrum can be written as
\begin{equation}
	\begin{aligned}
		\Pi(k_1,k_2,\omega)&=\dfrac{1}{(2\gpi)^2}\int_{-\infty}^{\infty}\int_{-\infty}^{\infty}\gamma(\xi,0,\omega)\mathrm{e}^{-\mathrm{i}\xi\omega/U_c(\omega)}G_{pp}(0,\eta,\omega)\mathrm{e}^{\mathrm{i}k_1\xi}\mathrm{e}^{\mathrm{i}k_2\eta}\mathrm{d}\xi\mathrm{d}\eta\\
		&=\phi_x(k_1,\omega)\phi_z(k_2,\omega),
	\end{aligned}
\end{equation}
where
\begin{equation}
	\phi_x(k_1,\omega) = \dfrac{1}{2\gpi}\int_{-\infty}^{\infty}\gamma(\xi,0,\omega)\mathrm{e}^{-\mathrm{i}\xi\omega/U_c(\omega)}\mathrm{e}^{\mathrm{i}k_1\xi}\mathrm{d}\xi,
\end{equation}
and
\begin{equation}
	\phi_z(k_2,\omega)=\dfrac{1}{2\gpi}\int_{-\infty}^{\infty}G_{pp}(0,\eta,\omega)\mathrm{e}^{\mathrm{i}k_2\eta}\mathrm{d}\eta.
\end{equation}
Here, $\phi_z(k_2,\omega)$ is the spanwise wavenumber-frequency spectrum while $\phi_x(k_1,\omega)$ denotes the effects of both the coherence decay and the convection of turbulent eddies. Thus, $S_{pp}$ can be shown to be
\begin{equation}
	S_{pp}(\boldsymbol{X}_0,\omega)=\left(\dfrac{\omega Y_0c}{4\gpi c_0S_0^2}\right)^22\gpi d\sum_{m=-\infty}^{\infty}\int_{-\infty}^{\infty}|\mathcal{L}(k_1,2m\gpi/\lambda),\omega|^2\phi_x(k_1,\omega)\mathrm{d}k_1\phi_z(2m\gpi/\lambda,\omega).
	\label{predequa2}
\end{equation}
Equation (\ref{predequa2}) shows that the form of $\phi_x$ plays an important role in estimating the integral and hence the far-field sound.

Under the frozen turbulence assumption, as discussed in $\S$\ref{ss2}, the streamwise coherence function is equal to one, and the convection velocity is assumed to be a constant value. This implies that all eddies convect at the same velocity while maintaining their coherence. As a result, the cross-spectral density is found to be
\begin{equation}
	\begin{aligned}
		G_{pp}(\xi,\eta,\omega)&=G_{pp}(0,\eta,\omega)\mathrm{e}^{-\mathrm{i}\omega\xi/U_c}.
	\end{aligned}
\end{equation}
 It follows that
 \begin{equation}
 	\phi_x(k_1,\omega)=\delta(k_1-\omega/U_c).
 	\label{frozpi}
 \end{equation}

Substituting (\ref{frozpi}) into (\ref{predequa2}), we recover
\begin{equation}
	S_{pp}(\boldsymbol{X}_0,\omega)=\left(\dfrac{\omega Y_0c}{4\gpi c_0S_0^2}\right)^22\gpi d\sum_{m=-\infty}^{\infty}|\mathcal{L}(\bar{k}_1,2m\gpi/\lambda,\omega)|^2\phi_z(2m\gpi/\lambda,\omega),
	\label{oriequ}
\end{equation}
where $\bar{k}_1=\omega/U_c$. Equation (\ref{oriequ}) is identical to (2.56) in \citet{lyu2016prediction}. In this case, a given frequency $\omega$ is assumed to be associated with a specific wavenumber $\bar{k}_1$ through the convection velocity $U_c$. Therefore, only the convection of eddies is considered, while the streamwise distortion is neglected. In terms of noise prediction models, the sound response function $\mathcal{L}$ sees only the value at the convective wavenumber $\bar{k}_1$. This simplification may be the reason why most analytical models overestimate the noise reduction of serrated trailing edges.

To account for the non-frozen nature of the turbulent boundary layer quantitatively, a coherence decay function is needed. We use
\begin{equation}
	\gamma(\xi,0,\omega)=\mathrm{e}^{-|\xi|/l_x(\omega)},
\end{equation}
 as informed by (\ref{gammax}). By employing this approximation, we can evaluate the integral in (\ref{predequa2}) and obtain
\begin{equation}
	\phi_x(k_1,\omega)=\dfrac{l_x(\omega)}{\gpi[1+(k_1-\omega/U_c(\omega))^2l_x^2(\omega)]}.
	\label{phiNonF}
\end{equation}

Equation (\ref{phiNonF}) incorporates the decay of streamwise coherence caused by the distortion of turbulent eddies as they convect downstream. It can be seen that
wavenumbers around $\omega/U_c(\omega)$ still play a significant role in shaping the spectrum, however, the contribution to the spectrum is not limited to a one-to-one correspondence between a given frequency and a specific wavenumber. This spreading phenomenon is a notable spectral characteristic of the wall pressure fluctuations beneath a turbulent boundary. In the following analysis, we adopt the notation $\tilde{k}_1(\omega)$ to represent the frequency-dependent convective wavenumber, i.e. $\tilde{k}_1(\omega)=\omega/U_c(\omega)$. This parameter represents the dominant wavenumber on the frequency of $\omega$.

To obtain the acoustic prediction, we need to evaluate the integral of $|\mathcal{L}|^2\phi_x$ over the streamwise wavenumber $k_1$. As mentioned at the beginning of this section, rather than relying on numerical techniques, we aim to obtain a compact estimation for the integral, enabling a convenient prediction and determining the physical impact of non-frozen turbulence on TE noise.

\subsection{Approximation of the model}\label{ss42}

According to the Mean Value Theorem for Integrals \citep{stewart2011calculus}, for a given frequency $\omega$ and spanwise mode $m$, there exists a characteristic wavenumber $k_{1,m}^\ast(\omega)$ such that
\begin{equation}
		\int_{-\infty}^{\infty}|\mathcal{L}(k_1,2m\gpi/\lambda,\omega)|^2\phi_x(k_1,\omega)\mathrm{d}k_1=|\mathcal{L}(k_{1,m}^\ast(\omega),2m\gpi/\lambda,\omega)|^2\int_{-\infty}^{\infty}\phi_x(k_1,\omega)\mathrm{d}k_1.
\label{MeanVal}
\end{equation}

Substituting the expression of $\phi_x$ given in (\ref{phiNonF}) into (\ref{MeanVal}) and recognizing that
\begin{equation}
    \int_{-\infty}^{\infty}\dfrac{l_x(\omega)}{\gpi[1+(k_1-\tilde{k}_1(\omega))^2l_x^2(\omega)]}\mathrm{d}k_1=1,
\end{equation}
 we have
 \begin{equation}
 		\int_{-\infty}^{\infty}|\mathcal{L}(k_1,2m\gpi/\lambda,\omega)|^2\phi_x(k_1,\omega)\mathrm{d}k_1=|\mathcal{L}(C_m(\omega)\tilde{k}_1(\omega),2m\gpi/\lambda,\omega)|^2.
 		\label{integL}
 \end{equation}

Substituting (\ref{integL}) into (\ref{predequa2}), we see that the resulting non-frozen model is identical to the frozen model apart from the introduction of a correction coefficient $C_m(\omega)=k_{1,m}^\ast(\omega)/\tilde{k}_1(\omega)$ to the dominant convection wavenumber. For a given frequency, by determining the correction coefficient $C_m$, the far-field noise can be calculated using essentially the same frozen prediction model. However, due to the complex nature of the response function, determining this coefficient is challenging. Therefore, in subsequent analysis, we aim to find a convenient approximation for $|\mathcal{L}|^2$. From the analysis of \citet{lyu2016prediction}, it is known that
\begin{equation}
	|\mathcal{L}|^2\sim O(|a_m|^2),
\end{equation}
where
\begin{equation}
	a_m=\dfrac{\mathrm{e}^{\mathrm{i}m\gpi/2}}{2}\mathrm{sinc}(k_1h-m\gpi/2)+\dfrac{\mathrm{e}^{-\mathrm{i}m\gpi/2}}{2}\mathrm{sinc}(k_1h+m\gpi/2).
\end{equation}

Examining the shape of $|a_m|^2$, we see that it can be very well approximated by
\begin{equation}
	H_m(k_1h)=\dfrac{1}{4}\left(\dfrac{1}{(k_1h+m\gpi/2)^2+\widetilde{m}}+\dfrac{1}{(k_1h-m\gpi/2)^2+\widetilde{m}}\right),
\end{equation}
where
\begin{equation}
	\widetilde{m}=1-\dfrac{1}{m\gpi+2}.
\end{equation}

It can be seen that $H_m$ is a purely algebraic function of $k_1h$. Therefore, the correction coefficient can be determined analytically by solving
\begin{equation}
		H_m(C_m\tilde{k}_1h)=\int_{-\infty}^{\infty}H_m(k_1h)\phi_x(k_1,\omega)\mathrm{d}k_1
		\label{HmInt}.
\end{equation}
The integral in (\ref{HmInt}) can be found analytically to be $I_m/4$, and the explicit form of $I_m$ is provided in Appendix \ref{appd}. Subsequently, the correction coefficient $C_m$ can be computed as 
\begin{equation}
	C_m=\sqrt{\dfrac{m^2\gpi^2}{4}-\widetilde{m}+\dfrac{1+\sqrt{1+m^2\gpi^2I_m-m^2\gpi^2\widetilde{m} I_m^2}}{I_m}}\bigg/\sigma_2, 
\end{equation}
where $\sigma_1=h/l_x$ and $\sigma_2=\tilde{k}_1h$. The correction coefficient $C_m$ is determined by $\sigma_1$ and $\sigma_2$ only.

With the introduction and evaluation of $C_m$, , the far-field noise spectrum is shown to be
\begin{equation}
	S_{pp}(\boldsymbol{X}_0,\omega)=\left(\dfrac{\omega Y_0c}{4\gpi c_0S_0^2}\right)^22\gpi d\sum_{m=-\infty}^{\infty}|\mathcal{L}(C_m(\omega)\tilde{k}_1(\omega),2m\gpi/\lambda,\omega)|^2\phi_z(2m\gpi/\lambda,\omega).
	\label{predfin}
\end{equation} 
Equation (\ref{predfin}) is purposely cast into the same form as the frozen model so that the effects of non-frozen turbulence can be accounted for conveniently by a single correction coefficient $C_m(\omega)$. In the following section, we will show the prediction results obtained using (\ref{predfin}), along with a discussion of the rationality behind the approximations used in this section.

\subsection{Prediction results and discussion}\label{ss43}

Using the new model that incorporates the impact of non-frozen turbulence, we can now predict the far-field noise. We apply the model to flat plates with wide and narrow serrations and use the similar parameters to those employed in the preceding numerical simulations. In particular, the Mach number is chosen as $M_0=0.03$, while the chord length of the flat plate is chosen as $c=1.12$ m. The streamwise correlation length $l_x(\omega)$ and the spanwise wavenumber-frequency spectrum $\phi_z(k_2,\omega)$ are both obtained from the numerical simulation. As shown in figure \ref{cohlen}, Smol'yakov's model better captures the frequency-dependent variation of the convection velocity, hence it is selected to compute the frequency-dependent convection velocity for the new model. \myadd{For the frozen model, the same computed spanwise wavenumber-frequency spectrum and a constant convection velocity $U_c=0.7U_0$ are used}. The non-dimensionalized form of the far-field power spectral density
\begin{equation}
	\Psi(\boldsymbol{X},\omega)=\dfrac{2\gpi S_{pp}(\boldsymbol{X},\omega)}{C_\ast(\rho_0v_\ast^2)^2(d/c_0)},                      
\end{equation}
is used to facilitate a direct comparison with results from frozen models, where $C_\ast\approx0.1553$ and $v_\ast\approx0.03U_0$. 

\begin{figure}
	\centering
	\includegraphics[scale=1]{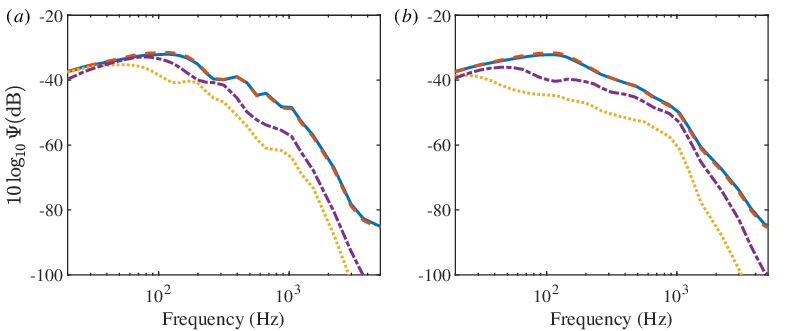}
	\caption{Predicted far-field noise produced by the straight and serrated trailing edges. The blue solid line represents the results for the straight trailing edge with a constant convection velocity. The orange dashed line corresponds to the straight trailing edge with the convection velocity calculated using the Smol'yakov model. The yellow dotted line corresponds to the serrated trailing edge using Lyu's original model. The purple dot-dashed line corresponds to the serrated trailing edge using the new model. (\textit{a}) $\lambda/h=2, h/c=0.025$ and (\textit{b}) $\lambda/h=0.4,h/c=0.05$.}
	\label{nonfropre}
\end{figure}

Figure \ref{nonfropre} presents the predicted far-field noise using both the frozen and non-frozen models. The observer is positioned at $90^\circ$ above the trailing edge and the distance between the observer and the trailing edge is equal to the chord length $c$. The wide serration has a size of $\lambda/h=2$ and $h/c=0.025$, while the narrow serration has a size of $\lambda/h=0.4$ and  $h/c=0.05$. For the straight trailing edge, results obtained using both a constant convection velocity $U_c=0.7U_0$ and the convection velocity calculated by the Smol'yakov model are shown. It can be seen that there are minimal discrepancies between the results obtained using these two types of convection velocities. Therefore, the frequency-dependent variation in convection velocity has a limited influence on the noise prediction for straight trailing edges. Significant noise reduction predicted by the frozen model can be seen within the intermediate- and high-frequency range for both wide and narrow serrations. Notably, narrow serrations perform better at intermediate frequencies, achieving noise reductions of over 15 dB. However, the new model that accounts for non-frozen turbulence predicts less significant noise reduction for both wide and narrow serrations. Moreover, the noise reduction appears less pronounced for narrow serrations compared to wide serrations. Such results have been reported in a number of experiments \citep{moreau2013noise,ning2017experimental,pereira2023parametric}. The long-standing discrepancies observed between the previous model and experiments may be explained if non-frozen turbulence is considered.

\begin{figure}
	\centering
	\includegraphics[scale=1]{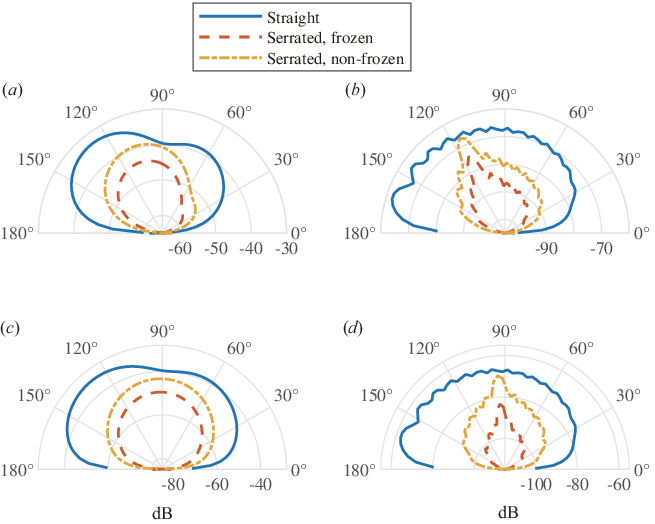}
	\caption{Directivity patterns for straight and serrated trailing edges predicted using frozen and non-frozen models: (\textit{a}) $\lambda/h=2, h/c=0.025$, $f=263$ Hz; (\textit{b}) $\lambda/h=2,h/c=0.025$, $f=2208$ Hz; (\textit{c}) $\lambda/h=0.4, h/c=0.05$, $f=263$ Hz and (\textit{d}) $\lambda/h=0.4, h/c=0.05$, $f=2208$ Hz.}
	\label{direc}
\end{figure}

The directivity patterns predicted by frozen and non-frozen models for different Mach numbers and frequencies are shown in figure \ref{direc}. The frozen model assumes a constant convection velocity of $U_c=0.7U_0$. It can be seen that the presence of serrated trailing edges significantly influences the directivity patterns, especially at higher frequencies (see figures \ref{direc}b and \ref{direc}d). Similar to those observed in \ref{nonfropre}, the new model predicts reduced levels of noise reduction for both wide and narrow serrations, while the shapes of the directivity patterns remain similar. From figure \ref{direc}, we can also see that the noise reduction effects are more pronounced in the regions located in front of and behind the flat plate.

In the prediction results shown above, the frequency-dependent correlation length and the spanwise wavenumber-frequency spectrum are obtained through numerical simulations. However, considering the high computational cost associated with computational fluid dynamics (CFD) or experimental investigations, the use of empirical models may be more convenient for practical applications. For example, we can use Chase's model to estimate the spanwise wavenumber-frequency spectrum, which is given by \citep{chase1987character,howe1991noise}
\begin{equation}
	\phi_z(k_2,\omega)\approx\dfrac{4C_\ast\rho_0^2v_\ast^4(\omega/U_c)^2\delta^4}{U_c\left(((\omega/U_c)^2+k_2^2)\delta^2+\chi^2\right)^2}.
\end{equation}
Here, $\chi\approx1.33$, and the boundary layer thickness $\delta$ is estimated using $\delta/c=0.382Re_c^{-1/5}$, where $Re_c$ represents the Reynolds number based on the chord length. In addition, the Smol'yakov model \citep{smol2006new} can be used to obtain the streamwise correlation length. Therefore, the new model demonstrates applicability across a broader range of scenarios.

\begin{figure}
	\centering
	\includegraphics[scale=1]{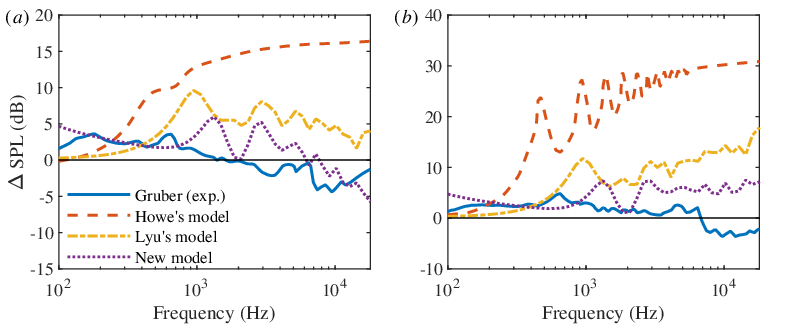}
	\caption{Comparison of the noise reduction predictions from analytical models with experimental measurements by \citet{gruber2012airfoil}. (\textit{a}) $\lambda/h=0.6$ and (\textit{b}) $\lambda/h=0.1$.}
	\label{ExpCom}
\end{figure}

To further investigate the accuracy of the new model, a comparison is conducted between the predicted noise reduction and that from experimental measurements by \citet{gruber2012airfoil}. Predictions from earlier models are also included for comparison. The experimental study used a NACA 65(12)-10 airfoil at a $5^\circ$ angle of attack with a free-stream velocity of $U_0=40$ $\mathrm{m/s}$. Two different sizes of serrations were used, namely $\lambda/h=0.6$ and $\lambda/h=0.1$. Due to the unavailability of the wall pressure statistics in the experiments, Chase's and the Smolyakov models are used in the prediction. 

The comparison results are shown in figure \ref{ExpCom}. In the experiments, noise reductions are observed at intermediate frequencies, albeit not exceeding approximately 5 dB. Conversely, noise increment appears at high frequencies. This phenomenon is consistent with the observations reported in the experiments conducted by \citet{oerlemans2009reduction}, which investigated full-scale serrated wind turbine blades. Howe's model exhibits a significant overprediction of the noise reduction, reaching a maximum $\Delta$SPL of $30$ dB for the narrow serrations (see figure \ref{ExpCom}b). On the other hand, Lyu's original model demonstrates a more realistic prediction, but overestimation is still pronounced. Comparatively, the discrepancies between the prediction using the new model and the experimental measurements are considerably smaller for both serrations. Therefore, we see that the frozen turbulence assumption contributes significantly to the overestimation of noise reduction. Interestingly, as shown in figure \ref{ExpCom}(a), noise increase at high frequencies is also predicted by the new model. This increment phenomenon has been attributed to the increased turbulent intensity between serration teeth \citep{gruber2012airfoil}. However, the present model suggests another possible cause of the noise increase at high frequencies. Figure \ref{ExpCom} clearly shows that it is necessary to account for the influence of non-frozen turbulence in acoustic predictions. 

\begin{figure}
	\centering
	\includegraphics[scale=0.9]{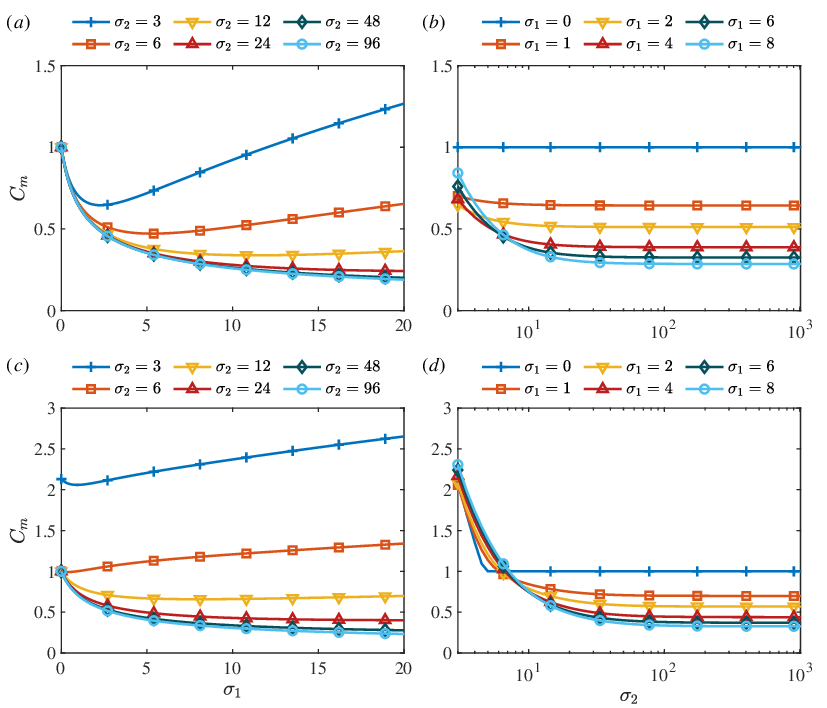}
	\caption{Correction coefficient $C_m$ as a function of $\sigma_1$ and $\sigma_2$: (\textit{a},\textit{b}) $m=0$ and (\textit{c},\textit{d}) $m=3$. Note the variations of $C_m$ are plotted from $\sigma_2=3$ in (\textit{b}) and (\textit{d}).}
	\label{corrCoeff}
\end{figure}  

\myadd{Equation (\ref{predfin}) shows that the effects of non-frozen turbulence are captured by a single coefficient $C_m(\omega)$. We can therefore study its variation and gain a better understanding of the consequence of including non-frozen turbulence. Figure \ref{corrCoeff} presents the variation of the correction coefficient $C_m$ as a function of $\sigma_1$ and $\sigma_2$. As shown in $\S$\ref{ss42}, $\sigma_1$ is defined as the ratio of the half amplitude to the frequency-dependent correlation length while $\sigma_2$ repsents the non-dimensional frequency. From figure \ref{corrCoeff}(a), we can see that for $m=0$, the correction coefficient $C_m$ initially decreases and then increases with the increase of $\sigma_1$ when $\sigma_2$ is small. However, when $\sigma_2$ attains large values, $C_m$ decreases monotonically. This implies that at high frequencies, more correctness is needed for the serrations with larger amplitudes. As shown in figure \ref{corrCoeff}(b), when $\sigma_1$ is set to 0, $C_m$ maintains a value of 1, indicating that no correction is needed for the straight trailing edge. When $\sigma_1$ is larger, $C_m$ initially decreases and then remains virtually constant. This suggests that, for serrated trailing edges, the correctness is nearly frequency-independent in the high-frequency range. For $m=3$, as shown in figures \ref{corrCoeff}(c) and \ref{corrCoeff}(d), $C_m$ continues to exhibit minor changes for large values of $\sigma_2$. However, the variation is more significant than $m=0$ when $\sigma_2$ is small.}  

\begin{figure}
	\centering
	\includegraphics[scale=0.88]{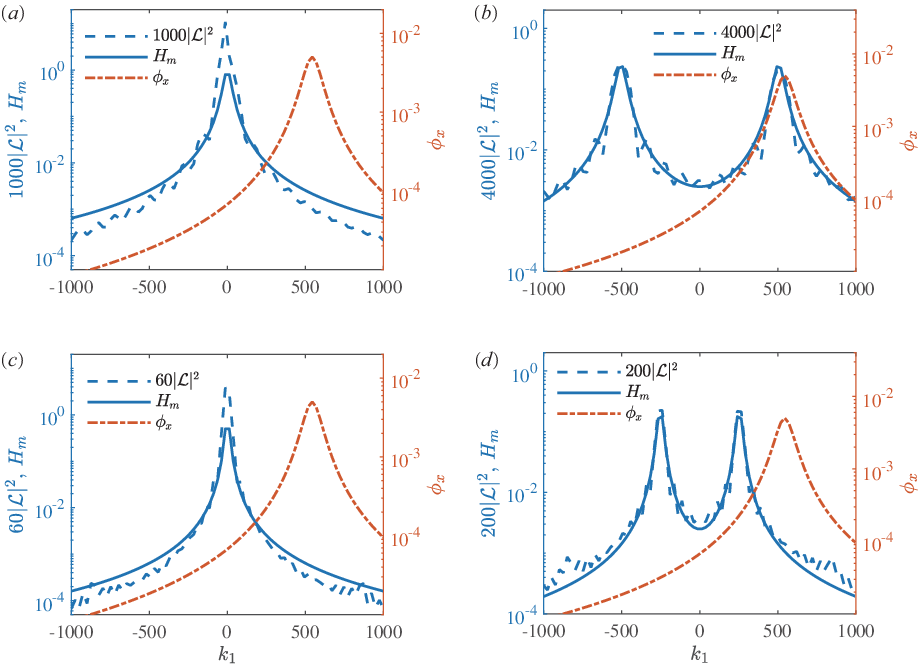}
	\caption{Comparison of the approximation functions $H_m$ and $|\mathcal{L}|^2$ multiplied by a constant value and the shape of $\phi_x$ at $600$ Hz with $M_0=0.03$: (\textit{a}) $m=0$, $\lambda/h=2$, $h/c=0.025$; (\textit{b}) $m=9$, $\lambda/h=2$, $h/c=0.025$; (\textit{c}) $m=0$, $\lambda/h=0.4$, $h/c=0.05$ and (\textit{d}) $m=9$, $\lambda/h=0.4$, $h/c=0.05$.}
	\label{approCom1}
\end{figure}

\begin{figure}
	\centering
	\includegraphics[scale=0.88]{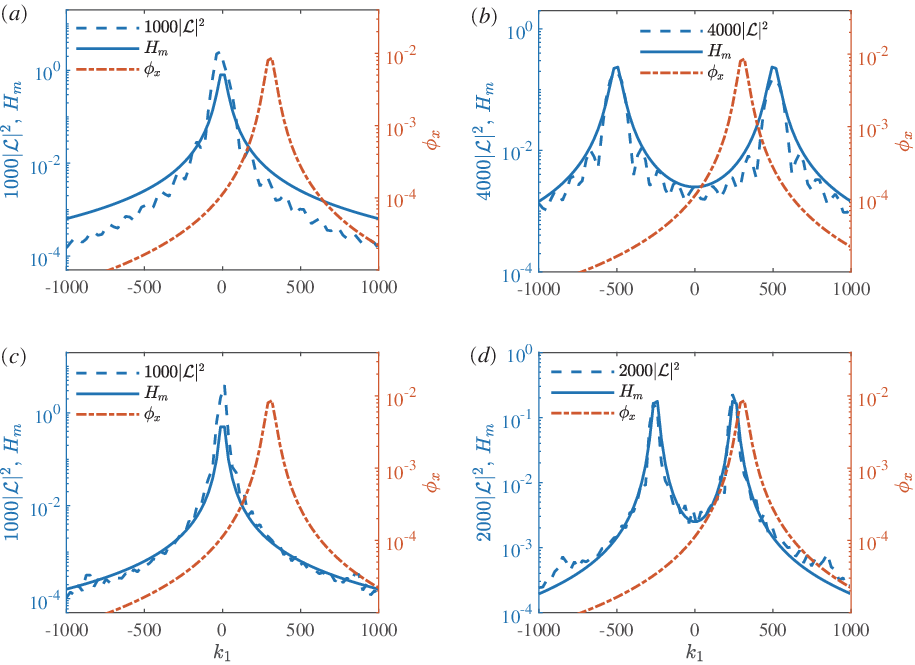}
	\caption{Comparison of the approximation functions $H_m$ and $|\mathcal{L}|^2$ multiplied by a constant value and the shape of $\phi_x$ at $2500$ Hz with $M_0=0.2$: (\textit{a}) $m=0$, $\lambda/h=2$, $h/c=0.025$; (\textit{b}) $m=9$, $\lambda/h=2$, $h/c=0.025$; (\textit{c}) $m=0$, $\lambda/h=0.4$, $h/c=0.05$ and (\textit{d}) $m=9$, $\lambda/h=0.4$, $h/c=0.05$.}
	\label{approCom2}
\end{figure}

To understand why the frozen turbulence assumption tends to overestimate noise reduction, we can study the integrand shown in (\ref{predequa2}) and its approximation in detail. To achieve that, the response function $|\mathcal{L}|^2$ and its approximation function $H_m$, as well as the shape of $\phi_x$ under non-frozen turbulence condition are shown in figures \ref{approCom1} and \ref{approCom2}. Note that the response function has been scaled by a constant factor for clearer comparison, without affecting the validity of the approximation. In addition, as pointed out by \citet{amiet1978noise}, the incorporation of the incident pressure raises the far-field sound by 6 dB. Here, the response function of only the scattered pressure is considered.   


In figure \ref{approCom1}, the comparison results at a frequency of $f=600$ Hz with a Mach number of $M_0=0.03$ are presented. It can be seen that the response function exhibits peaks near $k_1=0$ for $m=0$, but two peaks for $m=9$. The approximation function $H_m$ accurately captures the shape of the response function, particularly for $m=9$, indicating the validity of the approximation. 

As anticipated, it can be seen that the function $\phi_x$ exhibits a clear ridge at the streamwise wavenumber $\tilde{k}_1$. Since the calculation of far-field sound involves integrating the product of $|\mathcal{L}|^2$ and $\phi_x$ over the wavenumber $k_1$ (see (\ref{predequa2})), it is evident that the values near the peaks of both the response function and the convective ridge are important to the integral. However, assuming frozen turbulence results in the form of the Dirac delta function for $\phi_x$. Therefore, only the value of the response function at $\bar{k}_1$ is used, neglecting the influence of the shape of the response function. In contrast, when the turbulent boundary layer is not frozen, the shapes of both the response function and the convective ridge play important roles. For instance, as shown in figure \ref{approCom1}(a). The peaks of the response function and the convective ridge are sufficiently far away and the peaks near both $k_1=0$ and $k_1=\omega/U_c(\omega)$ contribute significantly to the integral. But if the single value $|\mathcal{L}(\bar{k}_1,k_2,\omega)|^2$ is used to represent the value of the integral as assumed by the frozen turbulence, the predicted far-field sound would become significantly lower than using the integral value. Conversely, if the peaks of the response function and the convective ridge are close to each other (see figure \ref{approCom1}b), the frozen turbulence assumption tends to predict a higher result. However, as shown in (\ref{predfin}), the overall far-field noise is the sum of contributions from all modes. It is known that the contribution of higher modes is less significant compared to lower modes. Therefore, for intermediate and high frequencies, where the peaks of the response function of the dominant modes and the convective ridge are not closely aligned, the analytical models assuming frozen turbulence would predict lower noise levels for serrated trailing edges. With regard to the impact of serration sizes, it can be seen from figures \ref{approCom1}(c) and \ref{approCom1}(d) that the shape of the response function is sharper for narrow serrations. Thus, more correctness to the convective wavenumber may be needed for narrow serrations.

To examine the effect of the Mach number, we present a similar comparison with a higher Mach number $M_0=0.2$ at a frequency of $f=2500$ Hz, as shown in figure \ref{approCom2}. It can be seen that with an increased Mach number, the convective ridge exhibits a sharper peak. Moreover, the convection velocity also increases with the increase of the Mach number. As shown in figure \ref{approCom2}(a), the peaks of the response function and the convective ridge become closer compared to the results shown in figure \ref{approCom1}, despite the frequency being increased from $600$ Hz to $2500$ Hz. This indicates that less correctness to the convective wavenumber may be needed for higher Mach numbers. The above analysis suggests that relying on the frozen turbulence assumption would result in more significant discrepancies for narrow serrations, especially at low Mach numbers.

In conclusion, we see that the frozen turbulence assumption may lead to lower or higher sound predictions for different spanwise modes, depending on the distance between the peaks of the response function and the convective ridge. However, when considering the collective contribution of all modes, lower noise levels would be predicted for serrated trailing edges in the intermediate- and high-frequency ranges. In other words, the frozen turbulence assumption results in an overestimated noise reduction. 

It is worth noting that the new model proposed in this study may have limited effectiveness when applied to very low frequencies. The first reason is that the phase velocity demonstrates significant variations within this frequency range, as shown in figure \ref{convel}(b). The second reason is that the exponential decay function assumed in the previous analysis may not accurately capture the behavior of the coherence loss. In this case, the introduction of a Gaussian phase decay term might be helpful to provide a more appropriate description of the coherence decay \citep{palumbo2013variance}. Nevertheless, in practical applications, it is the intermediate- and high-frequency ranges that are of most interest, because this is where noise reduction occurs. 

\section{Physical mechanism}\label{ss5}

In $\S$\ref{ss43}, a mathematical examination was conducted to explain the impact of the frozen turbulence assumption. In this section, we aim to elucidate the underlying physical mechanism related to noise reduction using non-frozen turbulence.

\begin{figure}
	\centering
	\includegraphics[scale=0.96]{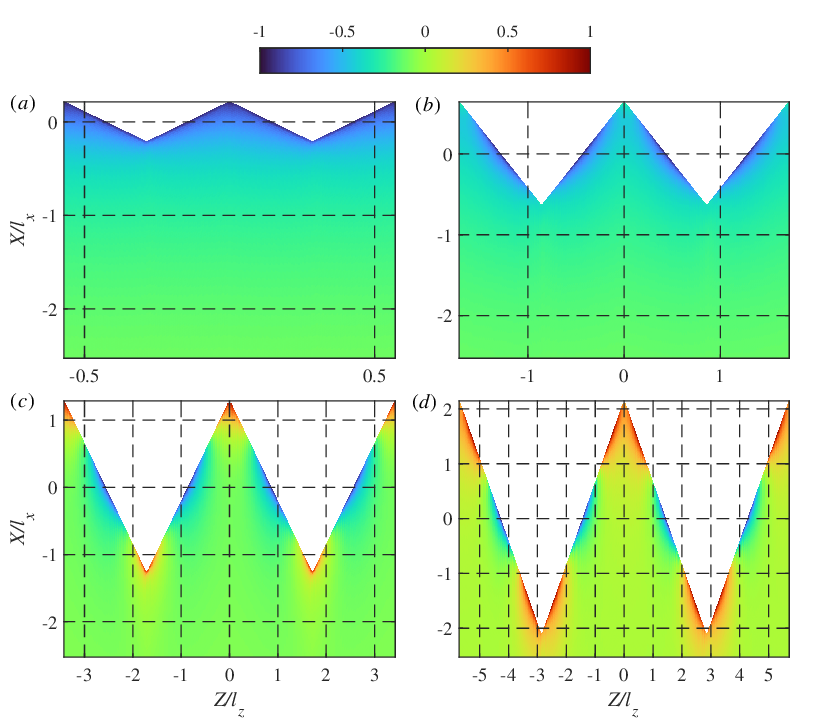}
	\caption{The scattered surface pressure distribution at a fixed frequency $f=2000$ Hz: (\textit{a}) $k_1h=2$, (\textit{b}) $k_1h=6$, (\textit{c}) $k_1h=12$ and (\textit{d}) $k_1h=20$. The distance between two adjacent dashed lines denotes the correlation lengths at this frequency.}
	\label{ScatPres}
\end{figure}

Previous work has shown that the physical mechanism behind the noise reduction can be attributed to the destructive interference of the scattered pressure \citep{lyu2016prediction}. To explore the efficiency of this interference, we show the scattered surface pressure distributions at a fixed frequency of $2000$ Hz with the Mach number $M_0=0.2$ in figure \ref{ScatPres}. The horizontal and vertical axes are scaled by the spanwise and streamwise frequency-dependent correlation lengths obtained using the Smol'yakov model, respectively. Hence, the distance between adjacent parallel dashed lines denotes the correlation length at this particular frequency.

As shown in figures \ref{ScatPres}(a) and \ref{ScatPres}(b), little phase variation appears when $k_1h$ attains a small value, indicating a weak noise reduction. However, as $k_1h$ increases, significant phase variation along the serration edges can be observed (see figures \ref{ScatPres}c and \ref{ScatPres}d). The interference resulting from this phase variation leads to noise reductions in the far field. Under the frozen turbulence assumption, the streamwise correlation length is assumed to be infinitely large. Consequently, the phase variations along the entire serration edges are considered to contribute to the destructive interference (assuming spanwise coherence is sufficiently large for now). However, in realistic non-frozen flows, only the phase variation within the streamwise correlation length is effective in the destructive interference. From figures \ref{ScatPres}(c) and \ref{ScatPres}(d), it can be seen that the serration amplitudes are 2-5 times larger than the streamwise correlation lengths, highlighting the significance of considering the streamwise length scale for accurate noise predictions.

In the above discussion, we purposely ignored the effects of spanwise correlation length for a simpler illustration. In realistic flows, both the spanwise and streamwise correlation lengths are important in determining the efficiency of the destructive interference. In fact, we can see from figure \ref{ScatPres} the 2D grids formed by the dashed lines that represent the streamwise and spanwise correlation lengths. It is within the same grid that the phase variation is effective. The 2D grid reflects the 3D structures of the turbulent flow. It can be seen that with the increase of $k_1h$, the grids become denser, indicating the effective area of destructive interference is smaller. Therefore the non-frozen correction must be included, particularly at high frequencies. The appearance of the 2D grid also explains why noise reduction does not increases monotonically with serration sharpness - the effective area of the destructive interference is heavily restricted by the spanwise and streamwise correlation lengths.  

\begin{figure}
	\centering
	\includegraphics[scale=1]{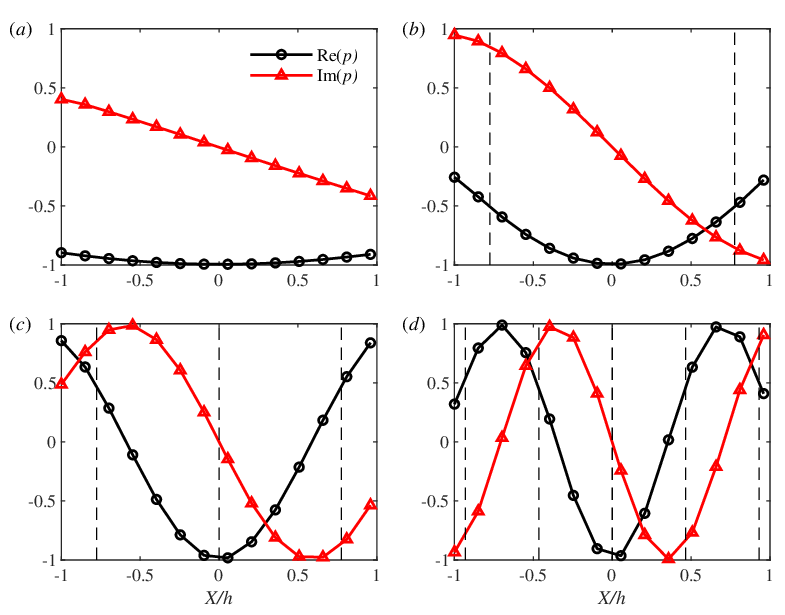}
	\caption{The real and imaginary parts of the scattered pressure on the serrated edge at a fixed frequency $f=2000$ Hz. (\textit{a}) $k_1h=2$, (\textit{b}) $k_1h=6$, (\textit{c}) $k_1h=12$ and (\textit{d}) $k_1h=20$. The distance between two adjacent dashed lines denotes the streamwise correlation length at this frequency.}
	\label{ScatPresLin}
\end{figure}

To provide further clarity on the phase variation along the serration edges, we plot the real and imaginary parts of the scattered pressure for different values of $k_1h$ in figure \ref{ScatPresLin}. The streamwise coordinates are normalized by half the amplitude of the serration $h$. Similarly, the distance between adjacent dashed lines corresponds to the streamwise correlation length. It can be seen from figure \ref{ScatPresLin}(a) that the real part remains negative for $k_1h=2$ and the imaginary part exhibits a phase that changes sign over the serration edge. Although the correlation length is larger than the serration amplitude, the noise reduction effect is not significant due to the insignificant phase variations. In figure \ref{ScatPresLin}(b), for $k_1h=6$, the phase differences of the scattered pressure are more significant, especially for the imaginary part. When $k_1h$ becomes larger, as shown in figures \ref{ScatPresLin}(c) and \ref{ScatPresLin}(d), strong variation can be seen for both the real and imaginary parts along the serration edge. However, it can be seen that these variations are less pronounced within a streamwise correlation length. Therefore, the interference effectiveness is not as strong as that assumed by the frozen turbulence. This explains why the frozen turbulence assumption tends to overestimate the noise reduction when employed in noise prediction models.          

In the case of leading-edge (LE) noise problems, where the inflow is typically uniform, the turbulent upwash velocity spectra can be accurately captured by various models, such as the von K\'arm\'an spectrum model \citep{amiet1975acoustic,narayanan2015airfoil}. By assuming frozen turbulence, analytical prediction models can provide realistic results for the noise emitted from LE serrations \citep{lyu2017noise}. However, this is not the case for TE serrations. As the turbulent boundary layer develops on a flat plate or an airfoil, the turbulent eddies undergo severe distortions due to the strong shear stresses, leading to significant streamwise coherence decay. Therefore, the impact of non-frozen turbulence must be taken into account. As shown in this study, a finite streamwise correlation length is introduced into the noise prediction model, resulting in significantly improved prediction accuracy.

\section{Conclusion}

This study investigates the impact of non-frozen turbulence on the noise prediction model for serrated trailing edges by analyzing the statistical characteristics of wall pressure fluctuations. A fully-developed turbulent boundary layer is simulated using LES, with the turbulence at the inlet generated by DFM. The accuracy of the simulated mean flow statistics is validated against DNS and a previous study by \citet{wang2022influence}. The simulation results demonstrate that as the spatial separations increase, the streamwise-spanwise correlation contour changes from circular to oval. Additionally, the space-time correlation contour lines concentrate into a narrow band. The mean convection velocity increases with the increase of streamwise separation while the phase velocities for a fixed streamwise separation initially increase and then decay with increasing frequency. Coherence function contours for both streamwise and spanwise directions are presented. The variation of the streamwise frequency-dependent correlation length indicates that the infinite streamwise correlation length assumed by frozen turbulence is not appropriate.

Lyu's model for serrated trailing edges is used as the basis for developing a non-frozen noise prediction model. This model involves integrating the product of the response function and the wavenumber-frequency spectrum over the streamwise wavenumber. Based on the statistical analysis of wall pressure fluctuations, an exponential coherence decay function is assumed, departing from the constant value employed under the frozen turbulence assumption. By examining the properties of the response function, an approximation function is introduced, allowing for the inclusion of a correction coefficient to account for the impact of non-frozen turbulence. Two non-dimensional parameters are identified to be critical for the non-frozen correctness, i.e. $\sigma_1=h/l_x$ and $\sigma_2=\tilde{k}h$. The far-field sound spectra for different serration sizes demonstrate that the new model predicts lower noise reduction. Comparative analysis with the experimental measurements of \citet{gruber2012airfoil} demonstrates that the new model has significantly better prediction capability. Furthermore, the noise increase at high frequencies may also be captured by the new model, suggesting a new cause for the high-frequency noise increase observed in various experiments. Through an examination of the response function and the convective ridge, it is shown that the far-field noise depends on the relative positions of their peaks. 

The physical mechanism underlying the overprediction of noise models employing the frozen turbulence assumption is found to be an overestimated destructive interference of the scatted pressure. As the non-dimensional parameter $k_1h$ increases, the streamwise correlation length becomes shorter than the amplitude of the serration. Only the phase variations within a streamwise correlation length can result in effective destructive interference. Consequently, the far-field noise is larger compared to that predicted under the frozen turbulence assumption. \myadd{This highlights the importance of the non-dimensional parameter $h/l_x$ as a crucial factor in determining the efficiency of destructive interference along the serration edge.}

It should be noted that the installation of serrations may alter the flow field near the trailing edge. The spectral properties may also change near the serrations \citep{pereira2022physics}. The present model relaxes the frozen turbulence assumption but still assumes that turbulence is statistically homogeneous in the streamwise direction. In applications where aerodynamic loading is present, the streamwise inhomogeneity may also need to be considered. This will be studied in our future works.


\section*{Acknowledgment} 
The first author (H.T.) would like to express gratitude to Dr. Yi Wang at Huairou Laboratory for the fruitful discussions on numerical simulations. The authors also acknowledge the financial support from Laoshan Laboratory under grant number LSKJ202202000.

\section*{Declaration of interests} 
The authors report no conflict of interest.
    
\appendix

\section{Mesh convergence test} \label{appa}

\begin{table}
	\centering
	\begin{tabular}{ccccccc}
		              & $n_x$ & $n_y$ & $n_z$ \\ 
              Coarse & 1500 & 34 & 90 \\
		     Middle & 2000 & 45 & 120\\
              Fine &2500 & 56 & 150 \\
	\end{tabular}
	\caption{Mesh parameters for different mesh sizes.}
	\label{gridcomp}
\end{table}

\begin{figure}
	\centering
	\includegraphics[scale=1]{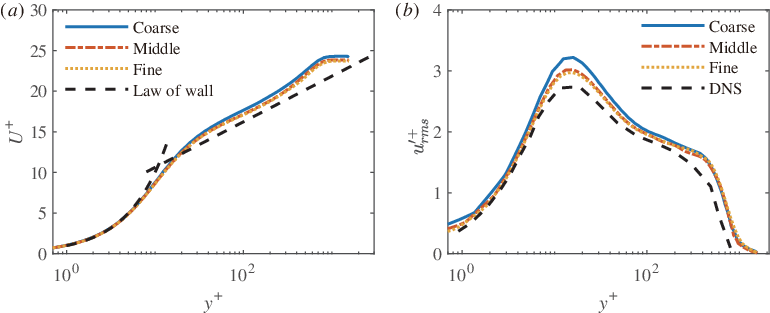}
	\caption{Comparison of the mean flow variables for the three meshes: (\textit{a}) mean velocity and (\textit{b}) streamwise fluctuating velocity.}
	\label{MeshComFig}
\end{figure}

To evaluate the convergence of the computational mesh, we use three different meshes: coarse, middle, and fine. These grid sizes are listed in Table \ref{gridcomp}. The resulting mean velocity and fluctuating streamwise velocity profiles, computed using these three meshes, are shown in figure \ref{MeshComFig}.  We can see the discrepancies between the results obtained with the fine and middle meshes were smaller compared to those between the middle and coarse meshes, particularly for the fluctuating streamwise velocity (see figure \ref{MeshComFig}b). Therefore, middle-size mesh is used in the simulation.

\section{Empirical convection velocity models}\label{appb}

The Bies model is given by \citep{bies1966review}
\begin{equation}
	\dfrac{U_c(\omega)}{U_0}=\left(\dfrac{U_0}{\omega\delta^\ast}\right)^{0.055}-0.3.
\end{equation}

The Smol'yakov model is given by \citep{smol2006new}
\begin{equation}
	\dfrac{U_c(\omega)}{U_0}=\dfrac{1.6\omega\delta^\ast/U_0}{1+16(\omega\delta^\ast/U_0)^2}+0.6.
	\label{UcSmol}
\end{equation}

\section{Empirical frequency-dependent correlation length models}\label{appc}

The Corcos model \citep{corcos1964structure} can be written as
\begin{equation}
	l_{x,z}(\omega)=\dfrac{U_c}{\alpha_{x,z}\omega},
\end{equation}
where $\alpha_x=0.11$, $\alpha_z=0.73$ and $U_c=0.7U_0$ are used.

The Smal'yakov model can be expressed as \citep{smol2006new}
\begin{equation}
	l_{x,z}(\omega)=\dfrac{U_c}{\alpha_{x,z}\omega}A^{-1},
\end{equation}
with
\begin{equation}
	A=\left[1-\dfrac{\beta U_c}{\omega\delta^\ast}+\left(\dfrac{\beta U_c}{\omega \delta^\ast}\right)^2\right]^{1/2}.
\end{equation}
Here, $\alpha_x=0.124$, $\alpha_z=0.8$ and $\beta=0.25$. The convection velocity is determined by employing (\ref{UcSmol}) and $\delta^\ast$ is approximated using $\delta^\ast/c\approx0.048/Re_c^{1/5}$.

Based on Goody's model \citep{goody2004empirical}, \citet{hu2021coherence} proposed an expression for the frequency-dependent correlation length, which can be written as
\begin{equation}
	\frac{l_x(\omega)}{\delta^*}=\frac{a\left(\omega \delta^* / U_0\right)^{0.3}}{(b+(\omega \delta^* / U_0)^{3.854})^{0.389}},
\end{equation}
with $a=1.357 \ln \left(Re_\theta\right)-6.713$, $b=1.183Re_\theta^{-0.593}$. And
\begin{equation}
	\frac{l_z(\omega)}{\delta^*}=\frac{a(\omega \delta^* / U_0)^{1.0}}{(b+(\omega \delta^* / U_0)^{3.073})^{0.651}},
\end{equation}
with $a=0.079 \ln \left(Re_\theta\right)+0.155$, $b=0.348Re_\theta^{-0.495}$.

\section{Expression of the integrated value $I_m$}\label{appd}

The expression of the integral result shown in $\S$\ref{ss41} is written as
\begin{equation}
	\begin{aligned}
		I_m=\dfrac{I_{m1}+I_{m2}}{I_{m3}+I_{m4}}+\dfrac{I_{m5}+I_{m6}}{I_{m7}+I_{m8}},
	\end{aligned}
\end{equation}
where
\begin{equation}
	I_{m1}=4\sqrt{\widetilde{m}}(4\widetilde{m}+\gpi^2m^2+4\gpi m\sigma_2+4\sigma_2^2-4\sigma_1^2),
\end{equation}
\begin{equation}
	I_{m2}=4(-4\widetilde{m} \sigma_1+\gpi^2m^2\sigma_1+4\gpi m\sigma_1\sigma_2+4\sigma_1\sigma_2^2+4\sigma_1^3),
\end{equation}
\begin{equation}
	I_{m3}=\sqrt{\widetilde{m}}(16\widetilde{m}^2+8(\gpi m+2\sigma_2-2\sigma_1 )(\gpi m+2\sigma_2+2\sigma_1)\widetilde{m}+\gpi^4m^4+8\gpi^3\sigma_2 m^3),
\end{equation}
\begin{equation}
	I_{m4}=\sqrt{\widetilde{m}}(24\gpi^2m^2\sigma_2^2+8\gpi^2m^2\sigma_1^2+32\gpi m\sigma_2^3+32\gpi m\sigma_1^2\sigma_2+16\sigma_2^4+32\sigma_1^2\sigma_2^2+16\sigma_1^4),
\end{equation}
\begin{equation}
	I_{m5}=4\sqrt{\widetilde{m}}(4\widetilde{m}+\gpi^2m^2-4\gpi m\sigma_2+4\sigma_2^2-4\sigma_1^2),
\end{equation}
\begin{equation}
	I_{m6}=4(-4\widetilde{m} \sigma_1+\gpi^2m^2\sigma_1-4\gpi m\sigma_1\sigma_2+4\sigma_1\sigma_2^2+4\sigma_1^3),
\end{equation}
\begin{equation}
	I_{m7}=\sqrt{\widetilde{m}}(16\widetilde{m}^2+8(\gpi m-2\sigma_2-2\sigma_1 )(\gpi m-2\sigma_2+2\sigma_1)\widetilde{m}+\gpi^4m^4-8\gpi^3\sigma_2 m^3), 
\end{equation}
\begin{equation}
	I_{m8}=\sqrt{\widetilde{m}}(24\gpi^2m^2\sigma_2^2+8\gpi^2m^2\sigma_1^2-32\gpi m\sigma_2^3-32\gpi m\sigma_1^2\sigma_2+16\sigma_2^4+32\sigma_1^2\sigma_2^2+16\sigma_1^4).
\end{equation}

\bibliographystyle{jfm}
\bibliography{non-frozen}

\end{document}